\documentclass[12pt]{article}
\usepackage{scicite}
\usepackage{times}
\usepackage{amsmath}
\usepackage{graphicx}%
\usepackage{xcolor}
\usepackage[labelfont=bf]{caption}
\makeatletter
\renewcommand{\fnum@figure}{Fig. \thefigure}

\makeatother

\newenvironment{sciabstract}{%
\begin{quote} \bf}
{\end{quote}}

\title{Spin-filter tunneling detection of antiferromagnetic resonance with electrically-tunable damping}

\author
{Thow Min Jerald Cham$^{1}\dagger$, Daniel G. Chica,$^{2}$ Xiaoxi Huang,$^{1}$ Kenji Watanabe$^{3}$,\\ Takashi Taniguchi$^{4}$, Xavier Roy$^{2}$, Yunqiu Kelly Luo$^{1,5,6,7,8}$*, Daniel C. Ralph$^{1,8}$*\\
\normalsize{Affiliations: $^{1}$Department of Physics, Cornell University;}\\
\normalsize{Ithaca, NY 14853, USA.}\\
\normalsize{$^{2}$Department of Chemistry, Columbia University;}\\
\normalsize{New York, NY, USA.}\\
\normalsize{$^{3}$Research Center for Electronic and Optical Materials, National Institute for Materials Science;}\\
\normalsize{Tsukuba, Ibaraki 305-0044, Japan.}\\
\normalsize{$^{4}$Research Center for Materials Nanoarchitectonics, National Institute for Materials Science;}\\
\normalsize{Tsukuba, Ibaraki 305-0044, Japan.}\\
\normalsize{$^{5}$Department of Physics and Astronomy, University of Southern California;}\\
\normalsize{Los Angeles, CA 90089, USA.}\\
\normalsize{$^{6}$Mork Family Department of Chemical Engineering and Materials Science, University of Southern}\\
\normalsize{California; Los Angeles, CA 90089, USA.}\\
\normalsize{$^{7}$Department of Chemistry, University of Southern California;}\\
\normalsize{Los Angeles, CA 90089, USA.}\\
\normalsize{$^{8}$Kavli Institute at Cornell;}\\
\normalsize{Ithaca, NY 14853, USA.}\\
\normalsize{*To whom correspondence should be addressed; E-mail: kelly.y.luo@usc.edu, dcr14@cornell.edu}\\
\normalsize{$\dagger$Present address: Department of Physics, California Institute of Technology; Pasadena,} \\
\normalsize{California 91125, USA.}
}

\date{}
\begin{document} 

\baselineskip24pt

\maketitle 

\begin{sciabstract}
Antiferromagnetic spintronics offers the potential for higher-frequency operations and improved insensitivity to magnetic fields compared to ferromagnetic spintronics. However, previous electrical techniques to detect antiferromagnetic dynamics have utilized large, millimeter-scale bulk crystals. Here we demonstrate direct electrical detection of antiferromagnetic resonance in structures on the few-micrometer scale using spin-filter tunneling in PtTe$_2$/bilayer CrSBr/graphite junctions in which the tunnel barrier is the van der Waals antiferromagnet CrSBr. This sample geometry allows not only efficient detection, but also electrical control of the antiferromagnetic resonance through spin-orbit torque from the PtTe electrode. The ability to efficiently detect and control antiferromagnetic resonance enables detailed studies of the physics governing these high frequency dynamics.
\end{sciabstract}

\noindent Manipulation of spin dynamics within antiferromagnets is attractive for future applications owing to the potential for high-frequency (GHz-THz) operation and insensitivity to small magnetic fields, but because antiferromagnets have no net magnetic moment, it is challenging to efficiently detect and control these dynamics\cite{jungwirth2016antiferromagnetic,RevModPhys.90.015005,han2023coherent}. The field of antiferromagnetic spintronics has made recent progress in demonstrating that the antiferromagnetic Néel vector can be reoriented using current pulses\cite{PhysRevLett.113.157201,science.aab1031,PhysRevLett.118.057701,bodnar2018writing,PhysRevApplied.9.064040, moriyama2018spin,PhysRevLett.120.207204,PhysRevLett.123.177201,PhysRevLett.129.017203} and in achieving large magnetoresistance in tunnel junctions made with metallic antiferromagnet electrodes\cite{qin2023room,chen2023octupole,shi2024electrically}.  High-frequency antiferromagnetic resonance modes have been detected using resonant absorption\cite{johnson1959antiferromagnetic,macneill2019gigahertz,Cham2022}, optical\cite{zhang2020gate,bae2022exciton,diederich2023tunable,sun2023universal}, and spin-pumping techniques\cite{li2020spin,vaidya2020subterahertz,wang2022spinpumping}. However, the previous electrical approaches of resonant absorption and spin pumping focused on millimeter-scale or larger samples. In order to utilize GHz-THz antiferromagnetic dynamics for applications such as radiation sources, modulators, and detectors, it will be necessary to develop much more compact electrical devices which are capable of both detecting and manipulating these dynamics in low-damping antiferromagnets. Here, we demonstrate micron-scale 3-terminal PtTe$_2$/bilayer CrSBr/graphite tunnel junctions which realize both functions. The devices achieve direct read-out of antiferromagnetic resonance in the CrSBr tunnel barrier using spin-filter tunneling\cite{worledge2000magnetoresistive,song2018giant,klein2018probing}, and at the same time allow the resonance damping to be tuned via spin-orbit torque\cite{ando2008electric,liu2011spin,miron2011perpendicular,liu2012spin} from the PtTe$_2$ electrode\cite{xu2020high,wang2024field}. 
The measurements reveal that the spin-orbit torque acts selectively only on the spin sublattice within the CrSBr layer adjacent to the PtTe$_2$ electrode.

\subsection*{3-terminal antiferromagnetic tunnel junction devices}
\noindent The antiferromagnet we employ, CrSBr\cite{Telford2020, lee2021magnetic,wilson2021interlayer,wu2022quasi,ziebel2024crsbr}, is an orthorhombic van der Waals (vdW) material in which spins within each layer are ferromagnetically aligned, whereas spins in adjacent layers are antiferromagnetically coupled.  The bulk N\'eel temperature is 132 K and the magnetic anisotropy is triaxial, with a hard axis along the out-of-plane crystal $\hat{c}$ axis, an easy axis along the crystal $\hat{b}$ axis, and an intermediate axis along the crystal $\hat{a}$ axis\cite{Telford2020,lee2021magnetic,wilson2021interlayer}. We will measure antiferromagnetic resonances for the case where a magnetic field is applied along or near the in-plane intermediate $\hat{a}$ axis, for which the equilibrium state of the two spin sublattices is a spin-flop configuration and the resonances have the form of in-phase (i.e., acoustic) or out-of-phase (i.e., optical) precession of the spin sublattices.  Antiferromagnetic resonances in CrSBr have been measured previously both by resonant absorption in bulk samples\cite{Cham2022} and by optical pump-probe methods in samples down to bilayer thicknesses\cite{bae2022exciton,diederich2023tunable}.

\vspace{0.5cm}
\noindent The spin-orbit-torque material we use, PtTe$_2$, is a vdW type-II Dirac semi-metal with an electrical conductivity of order 100 $\mu\Omega$cm and a spin-orbit torque efficiency per unit current density at room temperature of 0.05-0.15, comparable to Pt\cite{xu2020high,wang2024field}.

\vspace{0.5cm}
\noindent Our device geometry is shown in Fig.\ \ref{Figure1}A and B (see also Section 1.2 of \cite{SI}. It consists of a bottom PtTe$_2$ channel which makes contact to pre-formed Pt electrodes, a bilayer CrSBr flake on top of the PtTe$_2$ oriented so that the easy $\hat{b}$ axis of the crystal is 45$^\circ$ from the direction of current flow, and a narrow graphite top contact. The entire structure is encapsulated with a hexagonal boron nitride layer on top. All transport measurements to be reported in this paper were peformed in a liquid-nitrogen flow cryostat at 85 K, below the N\'eel temperature of CrSBr.

\vspace{0.5cm}
\noindent The low-bias magnetoresistance of the PtTe$_2$/bilayer CrSBr/graphite tunnel junction is shown in Fig.\ \ref{Figure2}A. The behavior differs in two critical ways from previous measurements which studied spin-filter tunneling between two graphite electrodes through antiferromagnetic CrI$_3$\cite{song2018giant,klein2018probing} or CrCl$_3$\cite{cai2019atomically}. First, the use of PtTe$_2$ instead of graphite for one of the electrodes reduces the overall tunnel-junction impedance, which is important for enabling high-frequency experiments. We measure a resistance of roughly 700 $\Omega$ for the PtTe$_2$/bilayer CrSBr/graphite device, whereas a comparable graphite/bilayer CrSBr/graphite device (not shown) had a resistance of 2000 $\Omega$. The second important difference is that CrSBr has significant within-plane magnetic anisotropy whereas CrI$_3$ and CrCl$_3$ have negligible anisotropy within their vdW planes. This magnetic anisotropy  causes the magnetoresistance of the CrSBr junction to differ for in-plane magnetic fields applied parallel and perpendicular to its easy axis.

\vspace{0.5cm}
\noindent When an applied magnetic field H is increased from zero in the direction parallel to the easy magnetic axis (the CrSBr $\hat{b}$ axis), we see an abrupt transition from a higher resistance into a lower resistance state at 0.1 Tesla that corresponds to a spin-flip transition of the CrSBr spin sublattices from an antiparallel state to a parallel state (Fig.\ \ref{Figure2}A). With H swept parallel to the intermediate anisotropy axis (the CrSBr $\hat{a}$ axis), there is instead a gradual transition from high to low resistance, corresponding to a gradual canting of both spin sublattices away from the easy axis. The ability to control the angle between the spin sublattices using a magnetic field applied along the intermediate axis allows us to maximize the sensitivity of the tunnel junctions for reading out the antiferromagnetic resonance and also to tune the magnetic damping using spin-orbit torque.

\subsection*{Spin-filter tunneling detection of antiferromagnetic resonance}
\noindent Antiferromagnetic resonance can be excited and detected electrically via a 3-terminal version of the spin-torque ferromagnetic resonance technique  \cite{xue2012resonance}, which for an antiferromagnetic resonance we will refer to as ST-AFMR, via the circuit diagram shown in Fig.\ \ref{Figure1}C. We apply a pulsed fixed-frequency microwave current (P $\leq$ 5 dBm) to the PtTe$_2$ channel through contact T1. Most of this current flows out through contact T2, but a small leakage current also flows through the tunnel junction to the top contact T3. The microwave current excites antiferromagnetic resonance through a combination of spin-orbit torque and the Oersted magnetic field, which results in an oscillating tunnel-junction resistance on account of spin-filter tunneling. Mixing between this oscillating resistance and an oscillating leakage current flowing through the tunnel junction results in a pulsed dc voltage at contact T3 that is measured using a lock-in amplifier. By sweeping an applied field through the resonant condition, the frequency of the resonance can be determined (Fig.\ \ref{Figure2}B). Figure\ \ref{Figure2}D shows the frequency \textit{vs.}\ field dependence for magnetic-field orientations near the intermediate anisotropy axis. In our devices, the leakage current through the tunnel junction is sufficient to produce a mixing signal with a large signal-to-noise ratio. If that had not been the case, we could also have applied a separate microwave current directly through the tunnel junction to achieve an even larger signal.

\vspace{0.5cm}
\noindent Of the two antiferromagnetic resonance modes, the measurement is sensitive only to the optical mode, because this is the mode in which the relative angle between the two spin sublattices undergoes large changes at the precession frequency to produce a substantial oscillating resistance.
Assuming a simple exchange field $\rm{H_E}$ between the spin sublattices in CrSBr, the field dependence for the optical-mode frequency for a magnetic field along the intermediate axis should have the form\cite{Cham2022}
\begin{align}
    {\rm{\omega_0}} &= \rm{\mu_0 \gamma \sqrt{\frac{((H_a + 2H_E)^2 - H^2)H_c}{H_a + 2H_E}}}
\label{eqn:omega2perp}
\end{align}
where $\rm{\mu_0}$ is the magnetic permeability, $\rm{\gamma}$ is the gyromagnetic ratio, H is the external field strength,  and $\rm{H_a}$ and $\rm{H_c}$ are the anisotropy parameters along the $\hat{a}$ and $\hat{c}$ axes respectively. From simultaneous fits to the resonance spectra in Fig.\ \ref{Figure2}D and the spin-flip transition in the tunneling magnetoresistance (Fig.\ \ref{Figure2}A), we obtain exchange and anisotropy parameter values $\rm{\mu_0 H_a}$ = 0.33(6) T, $\rm{\mu_0 H_E}$ = 0.096(1) T and $\rm{\mu_0 H_c}$ = 0.77(2) T (section 2.1 of \cite{SI}, Fig.\ S2, A and B). The in-plane and out-of-plane anisotropy parameters are not far from the values reported previously for bulk CrSBr at 85 K\cite{Cham2022} ($\rm{\mu_0 H_a^{bulk}} \approx$ 0.22, $\rm{\mu_0 H_c^{bulk}} \approx$ 0.75 T), but the exchange parameter is less than half the bulk value ($\rm{\mu_0 H_E^{bulk}}\approx$ 0.27 T). This may be caused by the reduced number of adjacent layers in a bilayer. As we rotate the field away from the intermediate anisotropy axis (Fig.\ \ref{Figure2}D) we see a scaling of the mode to smaller resonant fields, in agreement with previous measurements on bulk crystals\cite{Cham2022}.

\subsection*{Spin-orbit torque control of antiferromagnetic resonance linewidth}
\noindent Our 3-terminal device geometry allows for manipulation of the antiferromagnetic resonance in addition to simple detection of the resonance. If a direct charge current is applied within the PtTe$_2$ channel together with the microwave current, the dc current exerts an anti-damping or damping spin-orbit torque (depending on the sign of the direct current) on the CrSBr that tunes the resonance linewidth. In conventional ferromagnetic spin-orbit torque devices, the effectiveness of anti-damping torque for tuning the linewidth is proportional to $\cos\theta$, where $\theta$ is the angle between the precession axis of the magnetization and the anti-damping spin-orbit-torque vector $\hat{\sigma}$, which lies in-plane and perpendicular to the current\cite{liu2011spin, kasai2014modulation}. Next we analyze how the effects of anti-damping torque depend on the orientations of the two spin sublattices in antiferromagnetic CrSBr.

\vspace{0.5cm}
\noindent Figure \ref{Figure3} shows how a direct current affects the linewidths, for a microwave frequency of 13.65 GHz and a magnetic field sweep along the intermediate axis of CrSBr. The resonant magnetic field at this frequency causes the two spin sublattices to be canted at a relative angle of $\rm{{\phi} = 2 \arccos{(\sqrt{1 - (\frac{H}{2H_E + H_a})^2})} \approx 80^\circ}$\cite{Cham2022}. Because the CrSBr crystal is situated with the intermediate axis oriented approximately 45$^\circ$ from the direction of current in the PtTe$_2$ channel, this means that near the resonance one of the spin sublattices is oriented approximately parallel to the applied current and the other approximately perpendicular to the current and hence parallel or antiparallel to the spin-orbit torque vector, $\hat{\sigma}$ (see Fig.\ \ref{Figure3}, A and B). For the field direction depicted in Fig.\ \ref{Figure3}A, we find the results in Fig.\ \ref{Figure3}C: a negative dc current yields a linewidth significantly narrower than a positive current.  For the opposite sign of magnetic field (Fig.\ \ref{Figure3}B), there is negligible dependence of the linewidth on dc current (Fig.\ \ref{Figure3}D). We can quantify the linewidths by fitting each resonance to a sum of a symmetric and an antisymmetric Lorentzian\cite{xue2012resonance, liu2011spin}, with an additional linear term to account for a non-resonant background\cite{xue2012resonance}; we define $\Delta$ as the half width at half maximum of the Lorentzians. The overall dependences of $\Delta$ on current for the two signs of magnetic field are summarized in Fig.\ \ref{Figure3}E.

\vspace{0.5cm}
\noindent From these results we conclude that the anti-damping spin-orbit torque from the PtTe$_2$ layer acts selectively on one of the spin sublattices in the CrSBr, the sublattice adjacent to the PtTe$_2$ interface (we will call this spin sublattice 1).  For the field configuration corresponding to Fig.\ \ref{Figure3}A and C, spin sublattice 1 is aligned with the axis of the spin-orbit torque vector $\hat{\sigma}$, giving the maximum applied anti-damping torque with the maximum current-induced modulation of the linewidth. For the opposite sign of applied field (Fig.\ \ref{Figure3}B and D), spin sublattice 1 is parallel to the current channel and hence perpendicular to $\hat{\sigma}$, yielding negligible anti-damping torque.  Spin sublattice 2 in this second case is parallel to $\hat{\sigma}$ and hence in the orientation favorable to receive an anti-damping torque, but our measurements indicate that nevertheless little of the spin current from the PtTe$_2$ penetrates to this second layer so there is little effect on the overall linewidth. Thus, our technique allows an individual spin sublattice within an antiferromagnet to be addressed selectively.

\vspace{0.5cm}
\noindent We can model these effects quantitatively using a coupled two-lattice Landau-Lifshitz-Gilbert-Slonczewski (LLGS) equation\cite{slonczewski1996current} (section 2.3 of \cite{SI}). Assuming that the anti-damping spin-orbit torque acts only on sublattice 1, the dependence of the optical-resonance linewidth on the dc bias current  I$\rm{_{dc}}$ for magnetic fields applied along the intermediate axis should have the approximate form:
\begin{align}
    \rm{\Delta}
    &= \rm{\frac{2\pi f}{\gamma\omega_1}(\omega_2\alpha + \gamma \frac{\hbar}{2e}\frac{\xi_{SH}I_{dc}}{M_s W t_{m} t_{nm}} \cos{\rm{\theta_1}})}
\label{eqn:Delta}
\end{align}
where we define the expressions %frequencies (see S.I. section II)}
\begin{align}
\begin{split}
    \rm{\omega_1} &= \rm{2\gamma\mu_0\frac{H_0H_c}{(2H_E + H_a)}} \\  
    \rm{\omega_2} &= \rm{\gamma\mu_0\frac{(2H_E + H_a)(2H_E + H_a + H_c) - H_0^2}{(2H_E + H_a)}}. \\ 
\end{split}
\end{align}
In these equations, $\rm{\theta_1}$ is the angle between $\hat{\sigma}$ and spin-sublattice 1, $\rm{f}$ is the driving frequency, $\gamma$ is the electron gyromagnetic ratio, $\alpha$ is the intrinsic damping parameter, $\hbar$ is the reduced Planck's constant, $\xi_{SH}$ is the spin-Hall efficiency, e is the electron charge, $\rm{M_s}$ is the saturation magnetization, W = 3 $\mu$m is the width of the PtTe$_2$ channel, t$_{m}$ = 0.79 nm is the thickness of one CrSBr monolayer, t$_{nm}$ = 93 nm is the thickness of the PtTe$_2$ layer, and H$_0$ is the resonant magnetic field.

\vspace{0.5cm}
\noindent A linear fit to $\Delta$ against $\frac{2\pi f}{\gamma}\frac{\omega_2}{\omega_1}$ at zero bias gives $\alpha = 0.066(2)$ (Fig.\ S3C). Based on Eq.\ (\ref{eqn:Delta}), the slope of the current-modulated linewidth, normalizing for frequency and field dependencies, should have a cosine dependence on $\rm{\theta_1}$:
\begin{align}
\begin{split}
    \rm{\frac{d\Delta}{dI_{dc}}\frac{\omega_1}{2 \pi f}} &= \rm{\left(\frac{\hbar}{2e}\frac{\xi_{SH}}{M_s W t_{m} t_{nm}}\right)\cos{\rm{\theta_1}}}.
    \end{split}
     \label{eqn:cosprediction}
\end{align}
By changing the microwave frequency we can shift the resonance magnetic field and hence tune the canting angle of the spin sublattices near the resonance condition.  Figure \ref{Figure4}C shows how the current dependence of the damping of the optical mode depends on the orientation of spin sublattice 1 relative to the direction of the spin-orbit torque vector $\hat{\sigma}$ (i.e., the angles $\theta^{H-}_{1}$ and $\theta^{H+}_{1}$). For a magnetic field H applied along the intermediate anisotropy axis (i.e., 45$^\circ$ from $\hat{\sigma}$), we calculate these angles based on the previously-determined parameters $\rm{\mu_0H_E}$ = 0.096 T  and $\rm{\mu_oH_a}$ = 0.33 T\cite{Cham2022}:
\begin{equation}
\theta_1 = \begin{cases}
\rm{45^\circ - \sin^{-1}(H/(2H_E + H_a))} & \text{for } \rm{H < 0}\\
\rm{45^\circ + \sin^{-1}(H/(2H_E + H_a))} &\text{for } \rm{H > 0}.
\label{eqn:theta}
\end{cases}
\end{equation}
\noindent The measured dc-bias-modulated linewidth slopes at different angles in Fig.\ \ref{Figure4}C agree well with the $\cos{\rm{\theta_1}}$ dependence expected from Eq.\ ({\ref{eqn:cosprediction}), with no detectable dependence on $\theta_2$. We also verified Eq.\ (\ref{eqn:cosprediction}) by performing LLGS numerical calculations as a function of H and I$_{dc}$ (Fig.\ \ref{Figure4}D, {Figs.\ S7, S8 \cite{SI}). Using an estimated saturation magnetization value for bilayer CrSBr at 85 K of $\rm{\mu_0 M_s} = 0.28$ T (section 2.7 of \cite{SI}, Fig.\ S3D)}, a fit of the measurements to Eq.\ (\ref{eqn:cosprediction}) yields a value for the  anti-damping spin-orbit torque efficiency of $\rm{\xi_{SH}} = 0.29(2)$, somewhat larger than previous room temperature measurements for PtTe$_2$ which ranged from 0.05 to 0.15\cite{xu2020high, wang2024field}. We suspect the reason for this larger value may be that $\rm{\mu_0 M_s}$ is reduced by heating to make $\rm{\xi_{SH}}$ appear larger and/or the torque efficiency of PtTe$_2$ may be higher near 85 K than at room temperature. The substantial anti-damping spin-orbit torque we measure acting on the fully-uncompensated CrSBr interface is in contrast to a very small value measured previously for a spin-compensated $\alpha$-Fe$_2$O$_3$ interface\cite{cogulu2022}.

\vspace{0.5cm}
\noindent It is possible to reverse the N\'eel vector hysteretically by applying a magnetic field larger than 0.1 T along the magnetic easy axis of the CrSBr.  This might be a consequence of a small asymmetry in the magnetizations of the two layers, perhaps caused by differences arising from the PtTe$_2$/CrSBr interface versus the CrSBr/graphite interface. After the initial N\'eel vector is reversed relative to the configuration for the data in Fig.\ \ref{Figure3}E, the asymmetry of the current-dependent damping with respect to the sign of the applied magnetic field is also reversed (Fig.\ \ref{Figure3}F), as expected based on our assertion that only the spin sublattice at the PtTe$_2$ interface is affected by the anti-damping spin-orbit torque.

\subsection*{Discussion and Outlook}
\noindent The ability to both detect and control antiferromagnetic resonance using a tunnel junction structure opens the door to both fundamental studies of antiferromagnetic dynamics and potential high-frequency applications. With spin-filter tunneling, we achieve sensitive detection of antiferromagnetic resonance in a compact device geometry. The measurements we present represent only a first step in exploring the physics of antiferromagnetic spin dynamics; for example, studies with applied magnetic field in other directions than along the high-symmetry intermediate anisotropy axis should allow examination of the acoustic mode in addition to the optical mode, control over the degree of hybridization between modes \cite{Cham2022}, and studies of angular momentum flow between the spin sublattices (which has been shown to be important in synthetic antiferromagnets)\cite{mittelstaedt2021}.  Another direction is to create antiferromagnetic nano-oscillators for use as high-frequency sources – this will require using anti-damping torque to drive the effective damping of the antiferromagnetic resonance to negative values \cite{kiselev2003,urazhdin2012,liuoscillator,cheng2016terahertz}. In our existing devices, we achieve a maximum damping reduction of $\approx$ 12\% at a current density of 10$^{10}$ A/m$^2$, beyond which heating limits further decrease. Negative effective damping might be achieved by 
reducing the thickness of the PtTe$_2$ layer to minimize heating,
optimizing the ratio between antiferromagnetic parameters $\rm{\frac{d\Delta}{dI}} \propto \rm{\frac{2H_E + H_a}{H_0H_c}}$, decreasing the intrinsic damping $\alpha$, patterning the oscillator into a nanowire\cite{krivorotov2014}, or applying spin-orbit torque to both the top and bottom interfaces of the antiferromagnet.

\bibliography{scibib}

\begin{thebibliography}{10}

\bibitem{jungwirth2016antiferromagnetic}
T.~Jungwirth, X.~Marti, P.~Wadley, J.~Wunderlich, Antiferromagnetic spintronics, {\it Nature Nanotechnology\/} {\bf 11}, 231 (2016).

\bibitem{RevModPhys.90.015005}
V.~Baltz, A.~Manchon, M.~Tsoi, T.~Moriyama, T.~Ono, Y.~Tserkovnyak, Antiferromagnetic spintronics, {\it Rev. Mod. Phys.\/} {\bf 90}, 015005 (2018).

\bibitem{han2023coherent}
J.~Han, R.~Cheng, L.~Liu, H.~Ohno, S.~Fukami, Coherent antiferromagnetic spintronics, {\it Nature Materials\/} {\bf 22}, 684 (2023).

\bibitem{PhysRevLett.113.157201}
J.~\ifmmode~\check{Z}\else \v{Z}\fi{}elezn\'y, H.~Gao, K.~V\'yborn\'y, J.~Zemen, J.~Ma\ifmmode~\check{s}\else \v{s}\fi{}ek, A.~Manchon, J.~Wunderlich, J.~Sinova, T.~Jungwirth, Relativistic {N}\'eel-order fields induced by electrical current in antiferromagnets, {\it Phys. Rev. Lett.\/} {\bf 113}, 157201 (2014).

\bibitem{science.aab1031}
P.~Wadley, B.~Howells, J.~Železný, C.~Andrews, V.~Hills, R.~P. Campion, V.~Novák, K.~Olejník, F.~Maccherozzi, S.~S. Dhesi, S.~Y. Martin, T.~Wagner, J.~Wunderlich, F.~Freimuth, Y.~Mokrousov, J.~Kuneš, J.~S. Chauhan, M.~J. Grzybowski, A.~W. Rushforth, K.~W. Edmonds, B.~L. Gallagher, T.~Jungwirth, Electrical switching of an antiferromagnet, {\it Science\/} {\bf 351}, 587 (2016).

\bibitem{PhysRevLett.118.057701}
M.~J. Grzybowski, P.~Wadley, K.~W. Edmonds, R.~Beardsley, V.~Hills, R.~P. Campion, B.~L. Gallagher, J.~S. Chauhan, V.~Novak, T.~Jungwirth, F.~Maccherozzi, S.~S. Dhesi, Imaging current-induced switching of antiferromagnetic domains in {CuMnAs}, {\it Phys. Rev. Lett.\/} {\bf 118}, 057701 (2017).

\bibitem{bodnar2018writing}
S.~Y. Bodnar, L.~{\v{S}}mejkal, I.~Turek, T.~Jungwirth, O.~Gomonay, J.~Sinova, A.~Sapozhnik, H.-J. Elmers, M.~Kl{\"a}ui, M.~Jourdan, Writing and reading antiferromagnetic {Mn$_2$Au} by {N}{\'e}el spin-orbit torques and large anisotropic magnetoresistance, {\it Nature Communications\/} {\bf 9}, 348 (2018).

\bibitem{PhysRevApplied.9.064040}
M.~Meinert, D.~Graulich, T.~Matalla-Wagner, Electrical switching of antiferromagnetic {Mn}$_{2}${Au} and the role of thermal activation, {\it Phys. Rev. Appl.\/} {\bf 9}, 064040 (2018).

\bibitem{moriyama2018spin}
T.~Moriyama, K.~Oda, T.~Ohkochi, M.~Kimata, T.~Ono, Spin torque control of antiferromagnetic moments in {NiO}, {\it Scientific Reports\/} {\bf 8}, 14167 (2018).

\bibitem{PhysRevLett.120.207204}
X.~Z. Chen, R.~Zarzuela, J.~Zhang, C.~Song, X.~F. Zhou, G.~Y. Shi, F.~Li, H.~A. Zhou, W.~J. Jiang, F.~Pan, Y.~Tserkovnyak, Antidamping-torque-induced switching in biaxial antiferromagnetic insulators, {\it Phys. Rev. Lett.\/} {\bf 120}, 207204 (2018).

\bibitem{PhysRevLett.123.177201}
L.~Baldrati, O.~Gomonay, A.~Ross, M.~Filianina, R.~Lebrun, R.~Ramos, C.~Leveille, F.~Fuhrmann, T.~R. Forrest, F.~Maccherozzi, S.~Valencia, F.~Kronast, E.~Saitoh, J.~Sinova, M.~Kl\"aui, Mechanism of {N}\'eel order switching in antiferromagnetic thin films revealed by magnetotransport and direct imaging, {\it Phys. Rev. Lett.\/} {\bf 123}, 177201 (2019).

\bibitem{PhysRevLett.129.017203}
P.~Zhang, C.-T. Chou, H.~Yun, B.~C. McGoldrick, J.~T. Hou, K.~A. Mkhoyan, L.~Liu, Control of {N}\'eel vector with spin-orbit torques in an antiferromagnetic insulator with tilted easy plane, {\it Phys. Rev. Lett.\/} {\bf 129}, 017203 (2022).

\bibitem{qin2023room}
P.~Qin, H.~Yan, X.~Wang, H.~Chen, Z.~Meng, J.~Dong, M.~Zhu, J.~Cai, Z.~Feng, X.~Zhou, L.~Liu, T.~Zhang, Z.~Zeng, J.~Zhang, C.~Jiang, Z.~Liu, Room-temperature magnetoresistance in an all-antiferromagnetic tunnel junction, {\it Nature\/} {\bf 613}, 485 (2023).

\bibitem{chen2023octupole}
X.~Chen, T.~Higo, K.~Tanaka, T.~Nomoto, H.~Tsai, H.~Idzuchi, M.~Shiga, S.~Sakamoto, R.~Ando, H.~Kosaki, T.~Matsuo, D.~Nishio-Hamane, R.~Arita, S.~Miwa, S.~Nakatsuji, Octupole-driven magnetoresistance in an antiferromagnetic tunnel junction, {\it Nature\/} {\bf 613}, 490 (2023).

\bibitem{shi2024electrically}
J.~Shi, S.~Arpaci, V.~Lopez-Dominguez, V.~K. Sangwan, F.~Mahfouzi, J.~Kim, J.~G. Athas, M.~Hamdi, C.~Aygen, H.~Arava, C.~Phatak, M.~Carpentieri, J.~S. Jiang, M.~A. Grayson, N.~Kioussis, G.~Finocchio, M.~C. Hersam, P.~Khalili~Amiri, Electrically controlled all-antiferromagnetic tunnel junctions on silicon with large room-temperature magnetoresistance, {\it Advanced Materials\/} {\bf 36}, 2312008 (2024).

\bibitem{johnson1959antiferromagnetic}
F.~M. Johnson, A.~H. Nethercot~Jr., Antiferromagnetic resonance in {MnF$_2$}, {\it Physical Review\/} {\bf 114}, 705 (1959).

\bibitem{macneill2019gigahertz}
D.~MacNeill, J.~T. Hou, D.~R. Klein, P.~Zhang, P.~Jarillo-Herrero, L.~Liu, Gigahertz frequency antiferromagnetic resonance and strong magnon-magnon coupling in the layered crystal {CrCl$_3$}, {\it Phys. Rev. Lett.\/} {\bf 123}, 047204 (2019).

\bibitem{Cham2022}
T.~M.~J. Cham, S.~Karimeddiny, A.~H. Dismukes, X.~Roy, D.~C. Ralph, Y.~K. Luo, Anisotropic gigahertz antiferromagnetic resonances of the easy-axis van der {W}aals antiferromagnet {CrSBr}, {\it Nano Letters\/} {\bf 22}, 6716 (2022).

\bibitem{zhang2020gate}
X.-X. Zhang, L.~Li, D.~Weber, J.~Goldberger, K.~F. Mak, J.~Shan, Gate-tunable spin waves in antiferromagnetic atomic bilayers, {\it Nature Materials\/} {\bf 19}, 838 (2020).

\bibitem{bae2022exciton}
Y.~J. Bae, J.~Wang, A.~Scheie, J.~Xu, D.~G. Chica, G.~M. Diederich, J.~Cenker, M.~E. Ziebel, Y.~Bai, H.~Ren, C.~R. Dean, M.~Delor, X.~Xu, X.~Roy, A.~D. Kent, X.~Zhu, Exciton-coupled coherent magnons in a 2{D} semiconductor, {\it Nature\/} {\bf 609}, 282 (2022).

\bibitem{diederich2023tunable}
G.~M. Diederich, J.~Cenker, Y.~Ren, J.~Fonseca, D.~G. Chica, Y.~J. Bae, X.~Zhu, X.~Roy, T.~Cao, D.~Xiao, X.~Xu, Tunable interaction between excitons and hybridized magnons in a layered semiconductor, {\it Nature Nanotechnology\/} {\bf 18}, 23 (2023).

\bibitem{sun2023universal}
Y.~Sun, F.~Meng, C.~Lee, A.~Soll, H.~Zhang, R.~Ramesh, J.~Yao, Z.~Sofer, J.~Orenstein, Dipolar spin wave packet transport in a van der {W}aals antiferromagnet, {\it Nature Physics\/} {\bf 20}, 794 (2024).

\bibitem{li2020spin}
J.~Li, C.~B. Wilson, R.~Cheng, M.~Lohmann, M.~Kavand, W.~Yuan, M.~Aldosary, N.~Agladze, P.~Wei, M.~S. Sherwin, J.~Shi, Spin current from sub-terahertz-generated antiferromagnetic magnons, {\it Nature\/} {\bf 578}, 70 (2020).

\bibitem{vaidya2020subterahertz}
P.~Vaidya, S.~A. Morley, J.~van Tol, Y.~Liu, R.~Cheng, A.~Brataas, D.~Lederman, E.~Del~Barco, Subterahertz spin pumping from an insulating antiferromagnet, {\it Science\/} {\bf 368}, 160 (2020).

\bibitem{wang2022spinpumping}
L.~Wang, Y.~Zhao, Q.~Zhang, J.~Xue, J.~Guo, Y.~Chen, Y.~Tian, S.~Yan, L.~Bai, M.~Harder, N\'eel vector driven spin current in a van der {W}aals antiferromagnetic insulator ({C}r{C}l$_3$)/heavy metal ({P}t) bilayer, {\it Phys. Rev. B\/} {\bf 106}, 024422 (2022).

\bibitem{worledge2000magnetoresistive}
D.~C. Worledge, T.~H. Geballe, Magnetoresistive double spin filter tunnel junction, {\it Journal of Applied Physics\/} {\bf 88}, 5277 (2000).

\bibitem{song2018giant}
T.~Song, X.~Cai, M.~W.-Y. Tu, X.~Zhang, B.~Huang, N.~P. Wilson, K.~L. Seyler, L.~Zhu, T.~Taniguchi, K.~Watanabe, M.~A. McGuire, D.~H. Cobden, D.~Xiao, W.~Yao, X.~Xu, Giant tunneling magnetoresistance in spin-filter van der {W}aals heterostructures, {\it Science\/} {\bf 360}, 1214 (2018).

\bibitem{klein2018probing}
D.~R. Klein, D.~MacNeill, J.~L. Lado, D.~Soriano, E.~Navarro-Moratalla, K.~Watanabe, T.~Taniguchi, S.~Manni, P.~Canfield, J.~Fernández-Rossier, P.~Jarillo-Herrero, Probing magnetism in 2{D} van der {W}aals crystalline insulators via electron tunneling, {\it Science\/} {\bf 360}, 1218 (2018).

\bibitem{ando2008electric}
K.~Ando, S.~Takahashi, K.~Harii, K.~Sasage, J.~Ieda, S.~Maekawa, E.~Saitoh, Electric manipulation of spin relaxation using the spin {H}all effect, {\it Phys. Rev. Lett.\/} {\bf 101}, 036601 (2008).

\bibitem{liu2011spin}
L.~Liu, T.~Moriyama, D.~C. Ralph, R.~A. Buhrman, Spin-torque ferromagnetic resonance induced by the spin {H}all effect, {\it Phys. Rev. Lett.\/} {\bf 106}, 036601 (2011).

\bibitem{miron2011perpendicular}
I.~M. Miron, K.~Garello, G.~Gaudin, P.-J. Zermatten, M.~V. Costache, S.~Auffret, S.~Bandiera, B.~Rodmacq, A.~Schuhl, P.~Gambardella, Perpendicular switching of a single ferromagnetic layer induced by in-plane current injection, {\it Nature\/} {\bf 476}, 189 (2011).

\bibitem{liu2012spin}
L.~Liu, C.-F. Pai, Y.~Li, H.~W. Tseng, D.~C. Ralph, R.~A. Buhrman, Spin-torque switching with the giant spin {H}all effect of tantalum, {\it Science\/} {\bf 336}, 555 (2012).

\bibitem{xu2020high}
H.~Xu, J.~Wei, H.~Zhou, J.~Feng, T.~Xu, H.~Du, C.~He, Y.~Huang, J.~Zhang, Y.~Liu, H.-C. Wu, C.~Guo, X.~Wang, Y.~Guang, H.~Wei, Y.~Peng, W.~Jiang, G.~Yu, X.~Han, High spin {H}all conductivity in large-area type-{II} {D}irac semimetal {PtTe$_2$}, {\it Advanced Materials\/} {\bf 32}, 2000513 (2020).

\bibitem{wang2024field}
F.~Wang, G.~Shi, K.-W. Kim, H.-J. Park, J.~G. Jang, H.~R. Tan, M.~Lin, Y.~Liu, T.~Kim, D.~Yang, S.~Zhao, K.~Lee, S.~Yang, A.~Soumyanarayanan, K.-J. Lee, H.~Yang, Field-free switching of perpendicular magnetization by two-dimensional {PtTe$_2$/WTe$_2$} van der {W}aals heterostructures with high spin {H}all conductivity, {\it Nature Materials\/} pp. 1--7 (2024).

\bibitem{Telford2020}
E.~J. Telford, A.~H. Dismukes, K.~Lee, M.~Cheng, A.~Wieteska, A.~K. Bartholomew, Y.-S. Chen, X.~Xu, A.~N. Pasupathy, X.~Zhu, C.~R. Dean, X.~Roy, Layered antiferromagnetism induces large negative magnetoresistance in the van der {W}aals semiconductor {CrSBr}, {\it Adv. Mater.\/} {\bf 32}, 2003240 (2020).

\bibitem{lee2021magnetic}
K.~Lee, A.~H. Dismukes, E.~J. Telford, R.~A. Wiscons, J.~Wang, X.~Xu, C.~Nuckolls, C.~R. Dean, X.~Roy, X.~Zhu, Magnetic order and symmetry in the 2{D} semiconductor {CrSBr}, {\it Nano Letters\/} {\bf 21}, 3511 (2021).

\bibitem{wilson2021interlayer}
N.~P. Wilson, K.~Lee, J.~Cenker, K.~Xie, A.~H. Dismukes, E.~J. Telford, J.~Fonseca, S.~Sivakumar, C.~Dean, T.~Cao, X.~Roy, X.~Xu, X.~Zhu, Interlayer electronic coupling on demand in a 2{D} magnetic semiconductor, {\it Nature Materials\/} {\bf 20}, 1657 (2021).

\bibitem{wu2022quasi}
F.~Wu, I.~Guti{\'e}rrez-Lezama, S.~A. L{\'o}pez-Paz, M.~Gibertini, K.~Watanabe, T.~Taniguchi, F.~O. von Rohr, N.~Ubrig, A.~F. Morpurgo, Quasi-1d electronic transport in a 2d magnetic semiconductor, {\it Advanced Materials\/} {\bf 34}, 2109759 (2022).

\bibitem{ziebel2024crsbr}
M.~E. Ziebel, M.~L. Feuer, J.~Cox, X.~Zhu, C.~R. Dean, X.~Roy, {CrSBr}: An air-stable, two-dimensional magnetic semiconductor, {\it Nano Letters\/} {\bf 24}, 4319 (2024).

\bibitem{SI}
See {S}upplementary {M}aterials. .

\bibitem{cai2019atomically}
X.~Cai, T.~Song, N.~P. Wilson, G.~Clark, M.~He, X.~Zhang, T.~Taniguchi, K.~Watanabe, W.~Yao, D.~Xiao, M.~A. McGuire, D.~H. Cobden, X.~Xu, Atomically thin {CrCl$_3$}: an in-plane layered antiferromagnetic insulator, {\it Nano Letters\/} {\bf 19}, 3993 (2019).

\bibitem{xue2012resonance}
L.~Xue, C.~Wang, Y.-T. Cui, L.~Liu, A.~Swander, J.~Z. Sun, R.~A. Buhrman, D.~C. Ralph, Resonance measurement of nonlocal spin torque in a three-terminal magnetic device, {\it Phys. Rev. Lett.\/} {\bf 108}, 147201 (2012).

\bibitem{kasai2014modulation}
S.~Kasai, K.~Kondou, H.~Sukegawa, S.~Mitani, K.~Tsukagoshi, Y.~Otani, Modulation of effective damping constant using spin {H}all effect, {\it Applied Physics Letters\/} {\bf 104}, 092408 (2014).

\bibitem{slonczewski1996current}
J.~C. Slonczewski, Current-driven excitation of magnetic multilayers, {\it Journal of Magnetism and Magnetic Materials\/} {\bf 159}, L1 (1996).

\bibitem{cogulu2022}
E.~Cogulu, H.~Zhang, N.~N. Statuto, Y.~Cheng, F.~Yang, R.~Cheng, A.~D. Kent, Quantifying spin-orbit torques in antiferromagnet–heavy-metal heterostructures, {\it Phys. Rev. Lett.\/} {\bf 128}, 247204 (2022).

\bibitem{mittelstaedt2021}
J.~Mittelstaedt, D.~C. Ralph, Resonant measurement of nonreorientable spin-orbit torque from a ferromagnetic source layer accounting for dynamic spin pumping, {\it Phys. Rev. Appl.\/} {\bf 16}, 024035 (2021).

\bibitem{kiselev2003}
S.~I. Kiselev, J.~C. Sankey, I.~N. Krivorotov, N.~C. Emley, R.~J. Schoelkopf, R.~A. Buhrman, D.~C. Ralph, Microwave oscillations of a nanomagnet driven by a spin-polarized current, {\it Nature\/} {\bf 425}, 380 (2003).

\bibitem{urazhdin2012}
V.~E. Demidov, S.~Urazhdin, H.~Ulrichs, V.~Tiberkevich, A.~Slavin, D.~Baither, G.~Schmitz, S.~O. Demokritov, Magnetic nano-oscillator driven by pure spin current, {\it Nature Materials\/} {\bf 11}, 1028 (2012).

\bibitem{liuoscillator}
L.~Liu, C.-F. Pai, D.~C. Ralph, R.~A. Buhrman, Magnetic oscillations driven by the spin {H}all effect in 3-terminal magnetic tunnel junction devices, {\it Phys. Rev. Lett.\/} {\bf 109}, 186602 (2012).

\bibitem{cheng2016terahertz}
R.~Cheng, D.~Xiao, A.~Brataas, Terahertz antiferromagnetic spin {H}all nano-oscillator, {\it Phys. Rev. Lett.\/} {\bf 116}, 207603 (2016).

\bibitem{krivorotov2014}
Z.~Duan, A.~Smith, L.~Yang, B.~Youngblood, J.~Lindner, V.~E. Demidov, S.~O. Demokritov, I.~N. Krivorotov, Nanowire spin torque oscillator driven by spin orbit torques, {\it Nature Communications\/} {\bf 5}, 5616 (2014).

\bibitem{Zenodo}
Data and {LLGS} simulation software are deposited in the {Z}enodo repository, https://zenodo.org/records/15168545; doi: 10.5281/zenodo.15168545 .

\bibitem{Beck1990}
J.~Beck, Über chalkogenidhalogenide des chroms synthese, kristallstruktur und magnetismus von chromsulfidbromid, {CrSBr}, {\it Z. Anorg. Allg. Chem.\/} {\bf 585}, 157 (1990).

\bibitem{Scheie2022}
A.~Scheie, M.~Ziebel, D.~G. Chica, Y.~J. Bae, X.~Wang, A.~I. Kolesnikov, X.~Zhu, X.~Roy, Spin waves and magnetic exchange {H}amiltonian in {CrSBr}, {\it Adv. Sci.\/} {\bf 9}, 2202467 (2022).

\bibitem{ponomarenko2011}
L.~A. Ponomarenko, A.~K. Geim, A.~A. Zhukov, R.~Jalil, S.~V. Morozov, K.~S. Novoselov, I.~V. Grigorieva, E.~H. Hill, V.~V. Cheianov, V.~I. Fal’ko, K.~Watanabe, T.~Taniguchi, R.~V. Gorbachev, Tunable metal--insulator transition in double-layer graphene heterostructures, {\it Nat. Phys.\/} {\bf 7}, 958 (2011).

\bibitem{geim2013}
A.~K. Geim, I.~V. Grigorieva, Van der {W}aals heterostructures, {\it Nature\/} {\bf 499}, 419 (2013).

\bibitem{telford2022coupling}
E.~J. Telford, A.~H. Dismukes, R.~L. Dudley, R.~A. Wiscons, K.~Lee, D.~G. Chica, M.~E. Ziebel, M.-G. Han, J.~Yu, S.~Shabani, A.~Scheie, K.~Watanabe, T.~Taniguchi, D.~Xiao, Y.~Zhu, A.~N. Pasupathy, C.~Nuckolls, X.~Zhu, C.~R. Dean, X.~Roy, Coupling between magnetic order and charge transport in a two-dimensional magnetic semiconductor, {\it Nature Materials\/} {\bf 21}, 754 (2022).

\bibitem{lopez2022dynamic}
S.~A. L{\'o}pez-Paz, Z.~Guguchia, V.~Y. Pomjakushin, C.~Witteveen, A.~Cervellino, H.~Luetkens, N.~Casati, A.~F. Morpurgo, F.~O. von Rohr, Dynamic magnetic crossover at the origin of the hidden-order in van der {W}aals antiferromagnet {CrSBr}, {\it Nature Communications\/} {\bf 13}, 4745 (2022).

\bibitem{tschudin2023nanoscale}
M.~A. Tschudin, D.~A. Broadway, P.~Siegwolf, C.~Schrader, E.~J. Telford, B.~Gross, J.~Cox, A.~E.~E. Dubois, D.~G. Chica, R.~Rama-Eiroa, E.~J.~G.~Santos, M.~Poggio, M.~E. Ziebel, C.~R. Dean, X.~Roy, P.~Maletinsky, Imaging nanomagnetism and magnetic phase transitions in atomically thin {CrSBr}, {\it Nature Communications\/} {\bf 15}, 6005 (2024).

\bibitem{momma2011vesta}
K.~Momma, F.~Izumi, Vesta 3 for three-dimensional visualization of crystal, volumetric and morphology data, {\it Journal of Applied Crystallography\/} {\bf 44}, 1272 (2011).

\bibitem{boix2022probing}
C.~Boix-Constant, S.~Ma{\~n}as-Valero, A.~M. Ruiz, A.~Rybakov, K.~A. Konieczny, S.~Pillet, J.~J. Baldov{\'\i}, E.~Coronado, Probing the spin dimensionality in single-layer {CrSBr} van der {W}aals heterostructures by magneto-transport measurements, {\it Advanced Materials\/} {\bf 34}, 2204940 (2022).

\bibitem{karimeddiny2021resolving}
S.~Karimeddiny, D.~C. Ralph, Resolving discrepancies in spin-torque ferromagnetic resonance measurements: {L}ineshape versus linewidth analyses, {\it Physical Review Applied\/} {\bf 15}, 064017 (2021).

\bibitem{Mittelstaedt2022}
J.~Mittelstaedt, {``Generation of Spin Currents in Ferromagnetic Materials"}, thesis, Cornell University (2022).

\end{thebibliography}
\bibliographystyle{Science}

\noindent \textbf{Acknowledgements:} We thank Maciej Olszewski and Aaron Windsor for experimental assistance, and Rakshit Jain, Youn Jue (Eunice) Bae, John Cenker, M\"{a}rta Tschudin, Kaifei Kang,  Kihong Lee, Kin Fai Mak, and Jie Shan for discussions. 

\vspace{0.5cm}
\noindent \textbf{Funding:} AFOSR/MURI project 2DMagic grant FA9550-19-1-0390 (TMJC)\\
US National Science Foundation grant DMR-2104268 (YKL)\\ 
Singapore Agency for Science, Technology, and Research (TMJC)\\  
Cornell Presidential Postdoctoral Fellowship (YKL)\\  
Center for Energy Efficient Magnonics, an Energy Frontier Research Center funded by the U.S. Department of Energy, Office of Science, Basic Energy Sciences at SLAC National Laboratory under contract DE-AC02-76SF00515 (XH)\\  
US National Science Foundation through the Cornell Center for Materials Research grant DMR-1719875 (shared facilities)\\ Cornell NanoScale Facility, a member of the National Nanotechnology Coordinated Infrastructure, supported by the US National Science Foundation through grant NNCI 2025233\\ 
Kavli Institute at Cornell\\  
US National Science Foundation through the Columbia MRSEC on Precision Assembled Quantum Materials (PAQM) award DMR-2011738 (DGC)\\
U.S. Department of Energy, Office of Basic Energy Sciences (DE-SC0025422) (YKL)\\
JSPS KAKENHI, grants 21H05233 and 23H02052 (KW,TT)\\ 
CREST grant JPMJCR24A5 (KW,TT)\\
Japan Science and Technology Agency (KW,TT)\\ 
World Premier International Research Center Initiative (WPI), MEXT, Japan (KW, TT)

\vspace{0.5cm}
\noindent \textbf{Author contributions:} Device fabrication and primary measurements: TMJC\\ 
Assistance with measurements: YKL, XH\\
LLGS analysis and micromagnetic simulations: TMJC\\
Crystal synthesis: DGC, supervised by XR\\
hBN: KW, TT\\
Supervision: DCR\\
Writing – original draft: TMJC, DCR\\
Writing – review $\&$ editing: all authors

\vspace{0.5cm}
\noindent \textbf{Competing interests:} The authors declare no competing interests.

\vspace{0.5cm}
\noindent \textbf{Data and materials availability:} Data and analysis programs are deposited in the Zenodo 
repository (52).

\section*{Supplementary Materials}
Materials and Methods\\
Supplementary Text\\
Figs.\ S1 to S12\\
References \textit{(53-63)}

\clearpage

\vfill
\begin{figure}[htpb]
    \centering
    \includegraphics[width=0.8\textwidth]{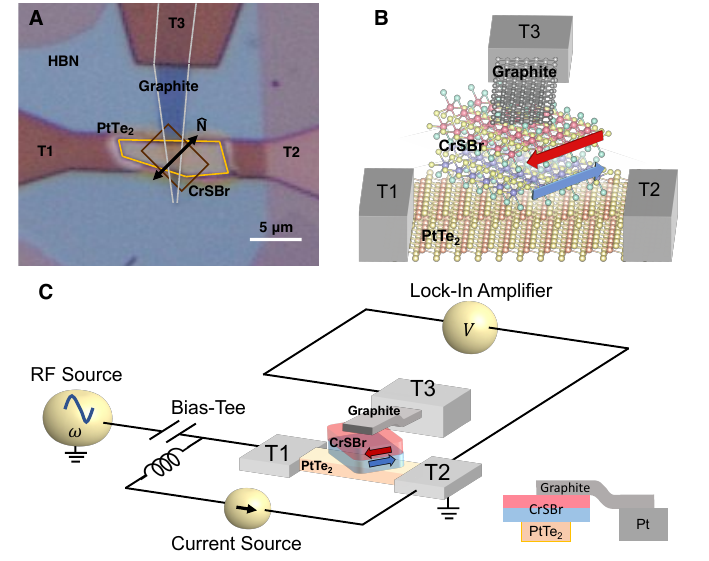}
        \caption{\textbf{PtTe$_2$/CrSBr/graphite 3-terminal device and measurement configuration.} (\textbf{A}) Optical microscopy image of the PtTe$_2$/CrSBr/graphite/hBN van der Waals heterostructure mechanically stacked and transferred onto Ti/Pt electrodes. $\hat{N}$ indicates the orientation of the N\'eel vector. (\textbf{B}) Schematic of the device structure (not to scale), showing alignment of the CrSBr antiferromagnetic sublattices (red and blue arrows) 45$^\circ$ with respect to the current in the PtTe$_2$ channel. (\textbf{C}) Schematic of the experimental circuit for ST-AFMR measurement of antiferromagnetic resonance.}\label{Figure1}
\end{figure}
\vfill

\null
\vfill
\begin{figure}[htpb]
    \centering
    \includegraphics[width=1.0\textwidth]{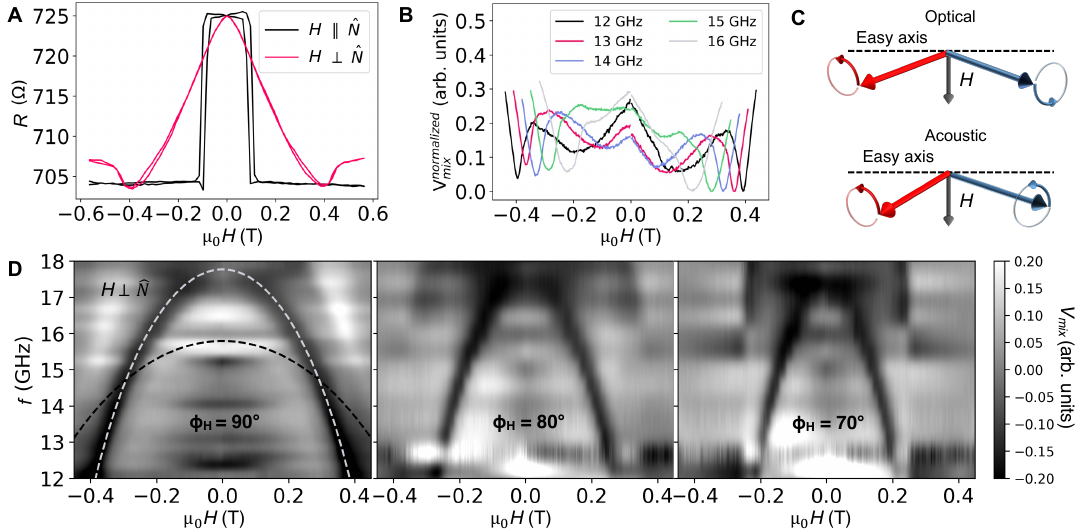}
    \caption{\textbf{Spin-filter tunneling magnetoresistance and ST-AFMR mixing voltage spectra.} (\textbf{A}) Magnetoresistance at 85 K measured between the top and bottom contacts (T3-T2 in Fig.\ 1) as a function of in-plane magnetic field H applied parallel to the CrSBr easy axis (H $\parallel \hat{N}$) and parallel to the intermediate axis (H $\perp \hat{N}$). (\textbf{B}) Antiferromagnetic resonances in the mixing voltage at 85 K as a function of magnetic field applied along the intermediate anisotropy axis, for frequencies 12-16 GHz. The mixing voltage curves shown here are normalized to account for frequency dependent transmission. (\textbf{C}) Depiction of out-of-phase optical and in-phase acoustic resonance modes. (\textbf{D}) ST-AFMR spectra of antiferromagnetic resonance for a magnetic field applied along the intermediate anisotropy axis ($\rm{\phi_H}$ = 90$^\circ$) and at two other nearby angles. In the $\rm{\phi_H}$ = 90$^\circ$ panel, the dashed gray line is a fit to the optical-magnon field dependence based on Eq.\ (\ref{eqn:omega2perp}) and the dashed black line is the corresponding field dependence expected for the acoustic mode, showing that the detection technique is not sensitive to the acoustic mode.}\label{Figure2}
\end{figure}
\vfill

\null
\vfill
\begin{figure}[htpb]
    \centering
    \includegraphics[width=0.9\textwidth]{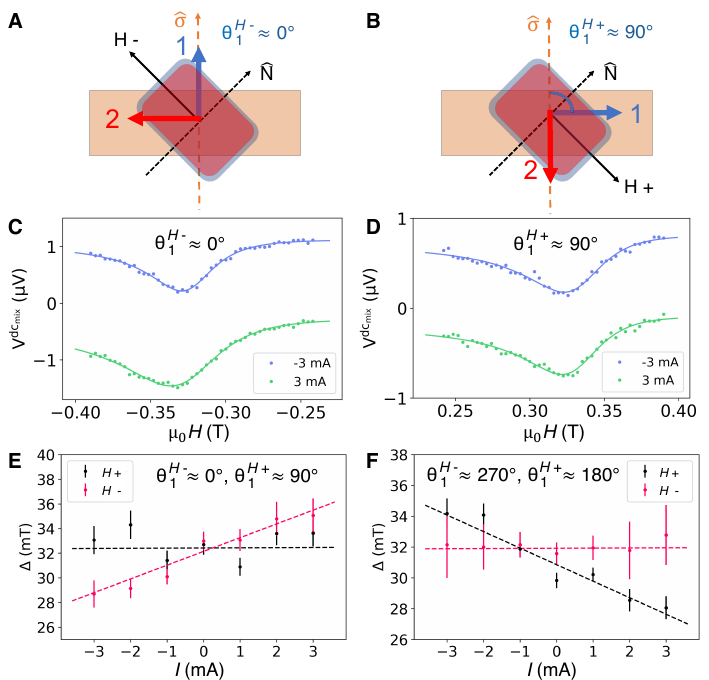}
    \caption{\textbf{ST-AFMR Measurements of resonant linewidth vs dc bias current at T = 85 K, f = 13.65 GHz.} (\textbf{A}, \textbf{B}) Schematic of the antiferromagnetic sublattice orientations for H- and H+ external fields, when sublattices 1 and 2 are canted approximately 90$^\circ$ apart. $\rm{\theta^{H-}_1}$ and $\rm{\theta^{H+}_1}$ are the relative angles between spin sublattice 1 and the spin-orbit-torque vector $\hat{\sigma}$ for negative and positive applied magnetic fields. (\textbf{C}, \textbf{D}) Rectified dc mixing voltage as a function of field at $\pm$ 3 mA for (\textbf{C}) negative and (\textbf{D}) positive external fields at f = 13.65 GHz. A non-resonant linear background was subtracted from the raw data (Fig. S12 \cite{SI}). (\textbf{E}) Resonant linewidth $\Delta$ vs dc bias current I$\rm{_{dc}}$ at f = 13.65 GHz, for negative (red) and positive (black) fields. (\textbf{F}) $\Delta$ vs I$\rm{_{dc}}$ after the sublattices have been reversed by a 0.4 T initialization field along the easy axis (see Fig.\ S8B for schematics of reversed configuration). Uncertainties in the linewidth $\Delta$ were determined from the standard deviation errors in the Lorentzian fits to the resonances.}
    \label{Figure3}
\end{figure}
\vfill

\newpage
\null
\vfill
\begin{figure}[htpb]
    \centering
    \includegraphics[width=0.9\textwidth]{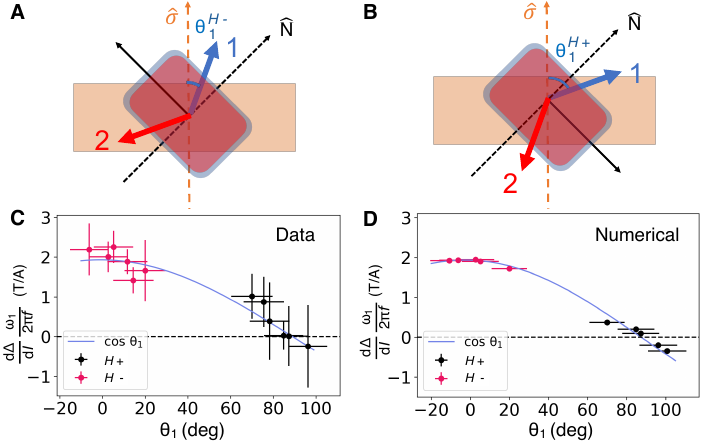}
    \caption{\textbf{Dependence of the dc-current-modulated linewidths on the orientation of spin sublattice 1.} (\textbf{A}, \textbf{B}) Illustration of the antiferromagnetic sublattices 1 and 2 canted by negative (H-) and positive (H+) fields along the intermediate anisotropy axis. $\rm{\theta^{H-}_1}$ and $\rm{\theta^{H+}_1}$ are the relative angles between spin sublattice 1 and the spin-orbit-torque vector $\hat{\sigma}$ for different values of H- and H+. (\textbf{C}) Slope of the modulated linewidth vs dc current for different orientations of spin sublattice 1, $\rm{\theta^{H-}_1}$ and $\rm{\theta^{H+}_1}$. (\textbf{D}) LLGS numerical calculations of dc current modulated linewidth slope for different configurations of $\rm{\theta^{H-}_1}$ and $\rm{\theta^{H+}_1}$. Uncertainties in the dc-bias modulated linewidth $\frac{d\Delta}{dI}$ were determined from the standard deviation errors of linear fits to $\Delta$ vs. I. Uncertainties in $\theta_1$ were determined from the width in magnetic field of the corresponding resonance.}
    \label{Figure4}
\end{figure} 
\vfill
{\color{white} \cite{Zenodo,Beck1990,Scheie2022,ponomarenko2011,geim2013,telford2022coupling,lopez2022dynamic,tschudin2023nanoscale,momma2011vesta,lee2021magnetic,Cham2022,boix2022probing,karimeddiny2021resolving,Mittelstaedt2022,macneill2019gigahertz,liu2011spin,xue2012resonance,ziebel2024crsbr}}
\end{document}

\baselineskip24pt
\title{Supplement to ``Spin-filter tunneling detection of antiferromagnetic resonance with electrically-tunable damping"
%“Electrical detection and damping control of antiferromagnetic resonance"
}
\author
{Thow Min Jerald Cham$^{1}$, Daniel G. Chica$^{2}$, Xiaoxi Huang$^{1}$, Kenji Watanabe$^{3}$,\\ Takashi Taniguchi$^{4}$, Xavier Roy$^{2}$, Yunqiu Kelly Luo$^{1,5,6\ast}$, Daniel C. Ralph$^{1,6\ast}$
\\
\normalsize{$^{1}$Department of Physics, Cornell University,}\\
\normalsize{Ithaca, NY 14853, USA}\\
\normalsize{$^{2}$Department of Chemistry, Columbia University,}\\
\normalsize{New York, NY, USA}\\
\normalsize{$^{3}$Research Center for Electronic and Optical Materials, National Institute for Materials Science}\\
\normalsize{Tsukuba, Ibaraki 305-0044, Japan}\\
\normalsize{$^{4}$Research Center for Materials Nanoarchitectonics, National Institute for Materials Science}\\
\normalsize{Tsukuba, Ibaraki 305-0044, Japan}\\
\normalsize{$^{5}$Department of Physics and Astronomy, University of Southern California}\\
\normalsize{Los Angeles, CA 90089, USA}\\
\normalsize{$^{6}$Kavli Institute at Cornell}\\
\normalsize{Ithaca, NY 14853, USA}\\
\normalsize{$^\ast$To whom correspondence should be addressed; E-mail: kelly.y.luo@usc.edu, dcr14@cornell.edu}
}

\maketitle
%\newpage
%\tableofcontents

\newpage
\section{Materials and Methods}
\subsection{CrSBr synthesis and heterostructure fabrication}
 \noindent CrSBr single crystals were synthesized using a modified chemical vapor transport approach originally adapted from Beck\cite{Beck1990}. Chromium, sulfur, and chromium tribromide reagents in a slightly off-stoichiometric ratio were sealed in a fused silica tube of 20 cm. The tube was heated in a two-zone tube furnace with a temperature gradient of 950 $^{\circ}$C to 850 $^{\circ}$C. Further details of the synthesis can be found in ref.\ \cite{Scheie2022}. Graphite and PtTe$_2$ single crystals were obtained commercially from HQGraphene.

\vspace{0.5cm}
 \noindent To fabricate the heterostructures, we exfoliated hBN, graphite, CrSBr and PtTe$_2$ flakes onto silicon/silicon dioxide (280 nm) wafers using the scotch-tape method. We selected flakes of appropriate thicknesses using an optical microscope equipped with a differential interference contrast prism for enhancing the optical contrast of steps in the flakes. We then mechanically picked up the selected flakes using stamps made from polypropylene carbonate (PPC) and polycarbonate (PC) and transferred onto pre-formed Pt contacts\cite{ponomarenko2011, geim2013}. During the pick-up procedure, we aligned the long edge of the CrSBr layer 45$^{\circ}$ to the bottom PtTe$_2$ channel. CrSBr flakes typically exfoliate into needle-shaped strips along the crystallographic $\hat{a}$ axis\cite{telford2022coupling,lopez2022dynamic,tschudin2023nanoscale}, which can be used as a reference for aligning the flakes during the stacking process. After completion of the transfer, we removed the polymer residue in chloroform, and rinsed the devices in acetone and then IPA before measurements. To prevent degradation, we stored the devices in a glovebox between measurements. Crystal structures of the vdW layers shown in Fig.\ 1 were drawn using VESTA\cite{momma2011vesta}.
 
\subsection{Device Details}
\begin{figure}[htpb]
    \centering
    \includegraphics[width=1.0\textwidth]{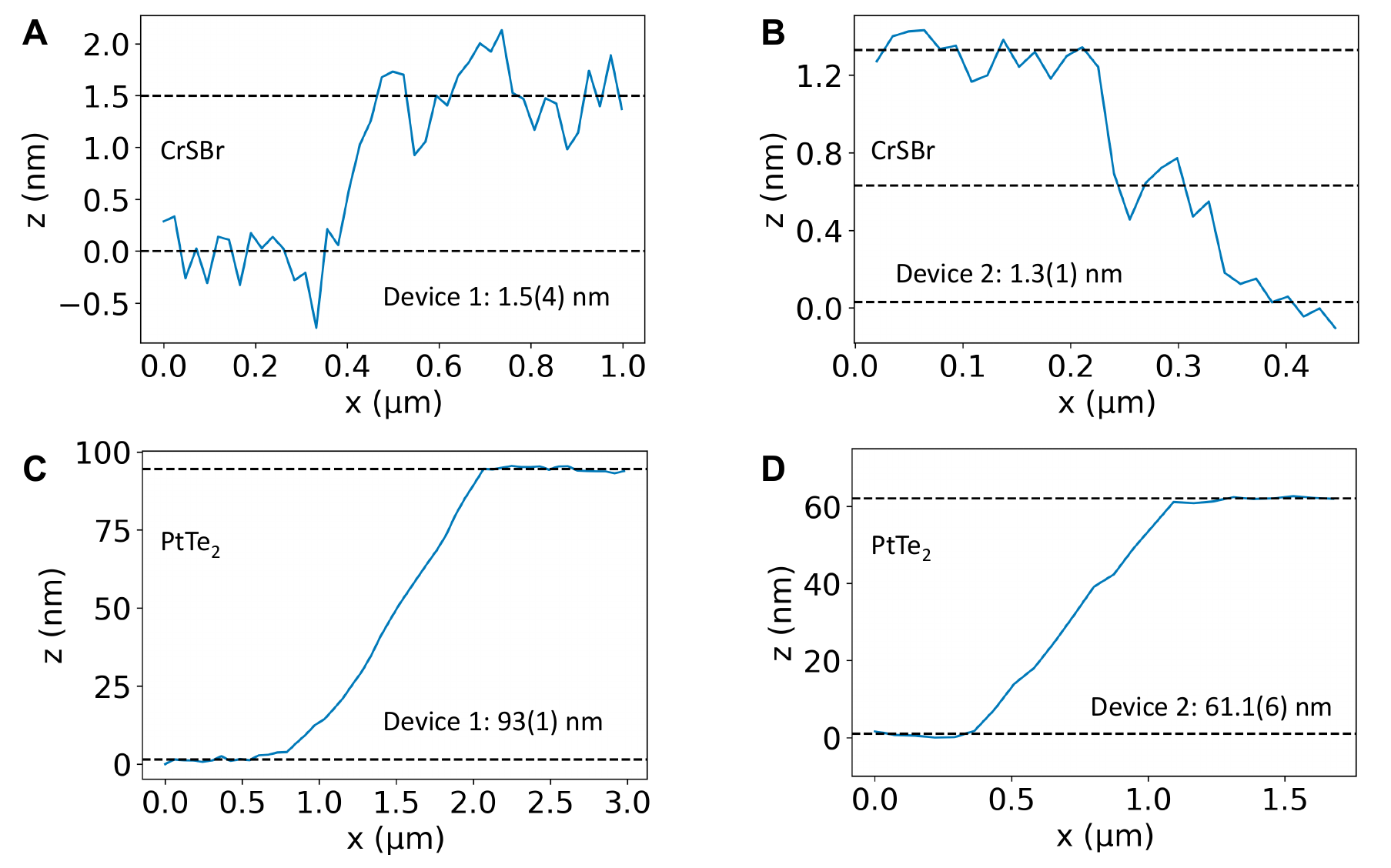}
    \caption{\textbf{Atomic force microscopy (AFM) linecuts of vdW layers in devices 1 and 2.} (\textbf{A}) AFM linecut of the CrSBr layer in device 1 with an estimated step height of t$\rm{_{CrSBr}}$ = 1.5(4) nm. (\textbf{B}) AFM linecut of the CrSBr layer in device 2 showing two steps at the flake edge, each with a size of t$\rm{_{CrSBr}}$ = 0.65(5), in approximate agreement with previous measurements of monolayer CrSBr t$\rm{_{CrSBr}}$ = 0.79(1)\cite{lee2021magnetic}. Noise in measurements is due to the thick hBN encapsulation layers ($\approx$ 20 nm) and polymer residue from the stacking process. (\textbf{C}) AFM linecut of the PtTe$_2$ layer in device 1 with an estimated step height of t$\rm{_{PtTe_2}}$ = 93(1) nm. (\textbf{D}) AFM linecut of the PtTe$_2$ layer in device 2 with an estimated step height of t$\rm{_{PtTe_2}}$ = 61.1(6) nm.}
    \label{FigureS1}
\end{figure}

 \noindent Based on atomic force microscope measurements (Fig.\ \ref{FigureS1}), the layer thicknesses for the device that is our focus in the main text are t$\rm{_{CrSBr}}$ = 1.5(4) nm, t$\rm{_{PtTe_2}}$ = 93(1) nm and t$\rm{_{graphite}}$ = 7(1) nm. The PtTe$_2$ channel resistance is 280 $\Omega$, and the area of the tunnel junction is approximately 5 $\rm{\mu m^2}$. Supplementary data for this device is shown in Figs.\ \ref{FigureS2}, \ref{FigureS3}, \ref{FigureS4}. Data from a second device are presented in Figs.\ \ref{FigureS5}, \ref{FigureS6}, \ref{FigureS7}, \ref{FigureS8}. Since the thickness of a monolayer of CrSBr is 0.79 nm\cite{lee2021magnetic}, the CrSBr tunnel barriers in both devices are bilayers. The CrSBr flakes in both samples are free of any monolayer steps within the device area. 

\subsection{Spin-torque ferromagnetic resonance}
\noindent The spin-orbit torque ferromagnetic resonance measurements were performed in a Janis flow cryostat with a maximum magnetic field of $\pm$0.6 T. An Agilent E8257D microwave frequency signal generator was used to apply the rf current, which was amplitude-modulated at a 1163.51 Hz. A signal recovery 7265 lock-in amplifier was used to measure the rectified mixing voltage.

\section{Supplementary Text}
\subsection{Magnetic parameters from fits to the resonance spectrum and tunneling magnetoresistance}

\begin{figure}[htpb]
    \centering
    \includegraphics[width=0.95\textwidth]{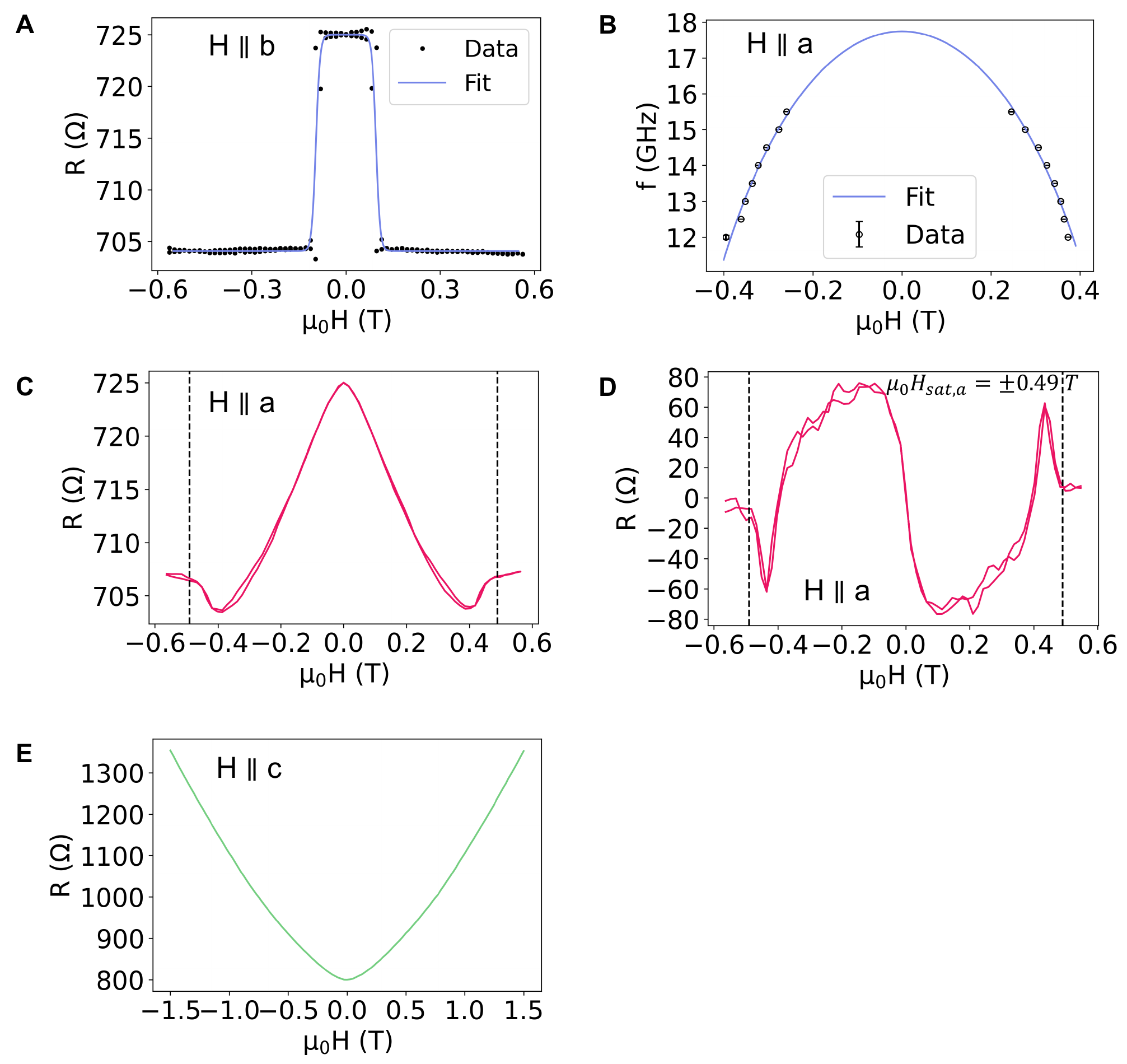}
    \caption{\textbf{Anisotropy and exchange parameters of device 1 analyzed in the main text.} (\textbf{A}, \textbf{B}) Simultaneous fits to the magnetoresistance and resonance spectrum of device 1 at 85 K using S.I.\ Eqs.\ S1 and S2, from which we extract the magnetic anisotropy and exchange parameters $\rm{\mu_0 H_a}$ = 0.33(6) T, $\rm{\mu_0 H_c}$ = 0.77(2) T, and $\rm{\mu_0 H_E}$ = 0.096(1) T. (\textbf{C}, \textbf{D}) Magnetoresistance and first derivative of the magnetoresistance as a function of external magnetic field along the $\hat{a}$ axis. Dashed lines indicate the estimated saturation magnetic field of $\mu_0H_{sat, a}$ $\approx$ 0.49 T. (\textbf{E}) Magnetoresistance as a function of external magnetic field along the $\hat{c}$ axis.}
    \label{FigureS2}
\end{figure}

\begin{figure}[htpb]
    \centering
    \includegraphics[width=0.9\textwidth]{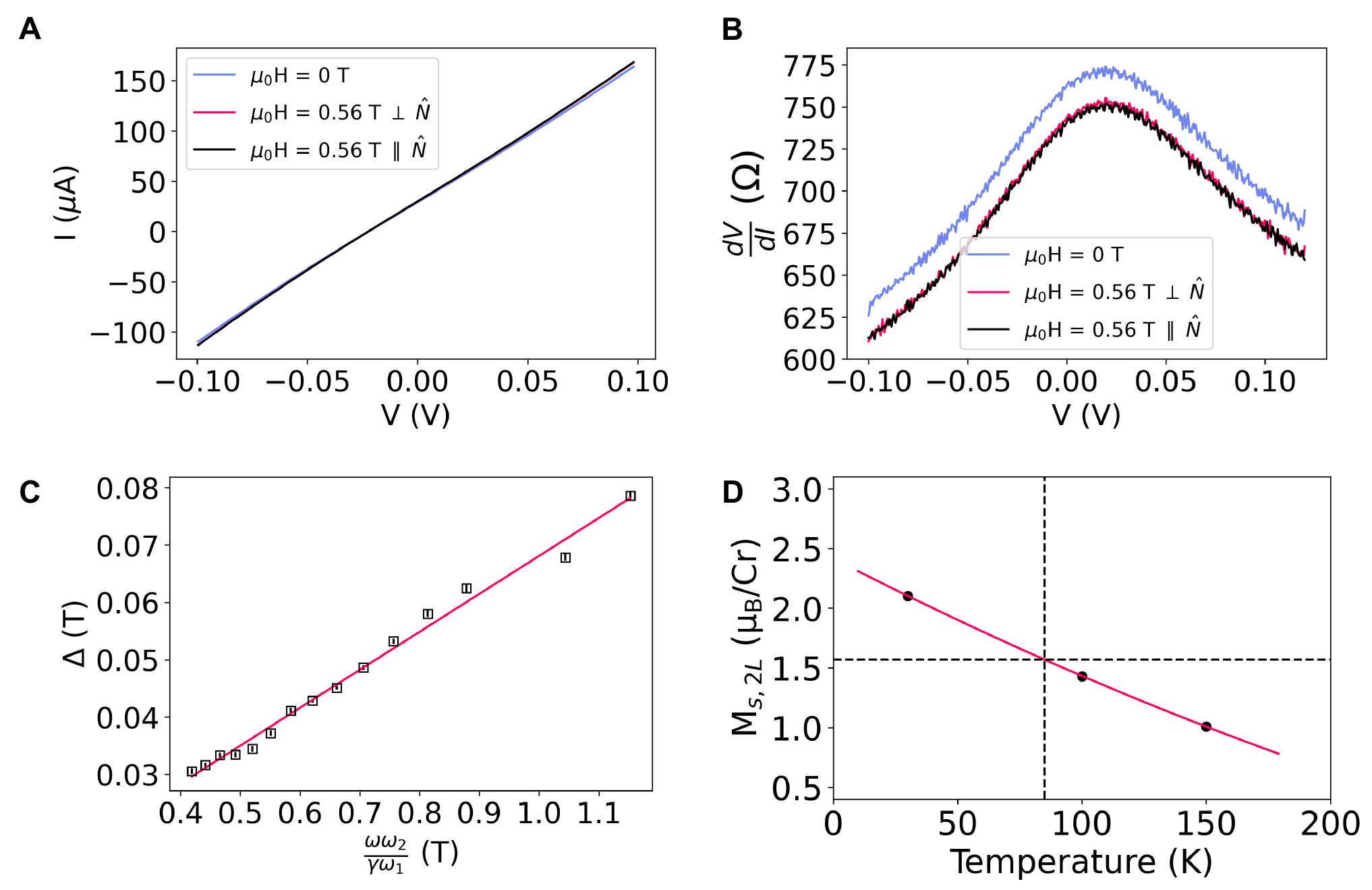}
    \caption{ \textbf{Supplementary characterization of device 1 analyzed in the main text.} (\textbf{A}, \textbf{B}) Transport measurements between T1 and T3 of device 1 at 85 K. (\textbf{A}) I-V curves at 0 field, H $\parallel\hat{N}$ and H $\perp\hat{N}$. (\textbf{B}) Corresponding differential resistance as a function of voltage. (\textbf{C}) Linear fit to resonance linewidth at different driving frequencies at 85 K, from which we extract a damping parameter $\alpha = 0.066(2)$ and an inhomogeneous linewidth broadening parameter $\Delta_0$ = 1.9(1.3) mT. (\textbf{D}) Saturation magnetization of bilayer CrSBr estimated from temperature series in Lee et al. \cite{lee2021magnetic}, thickness series in Tschudin et al.\cite{tschudin2023nanoscale} and assuming a bulk M$_s$ of 3 $\mu_B$ per Cr atom at 0 K\cite{lopez2022dynamic, ziebel2024crsbr}.}
    \label{FigureS3}
\end{figure}

\begin{figure}[htpb]
    \centering
    \includegraphics[width=1.0\textwidth]{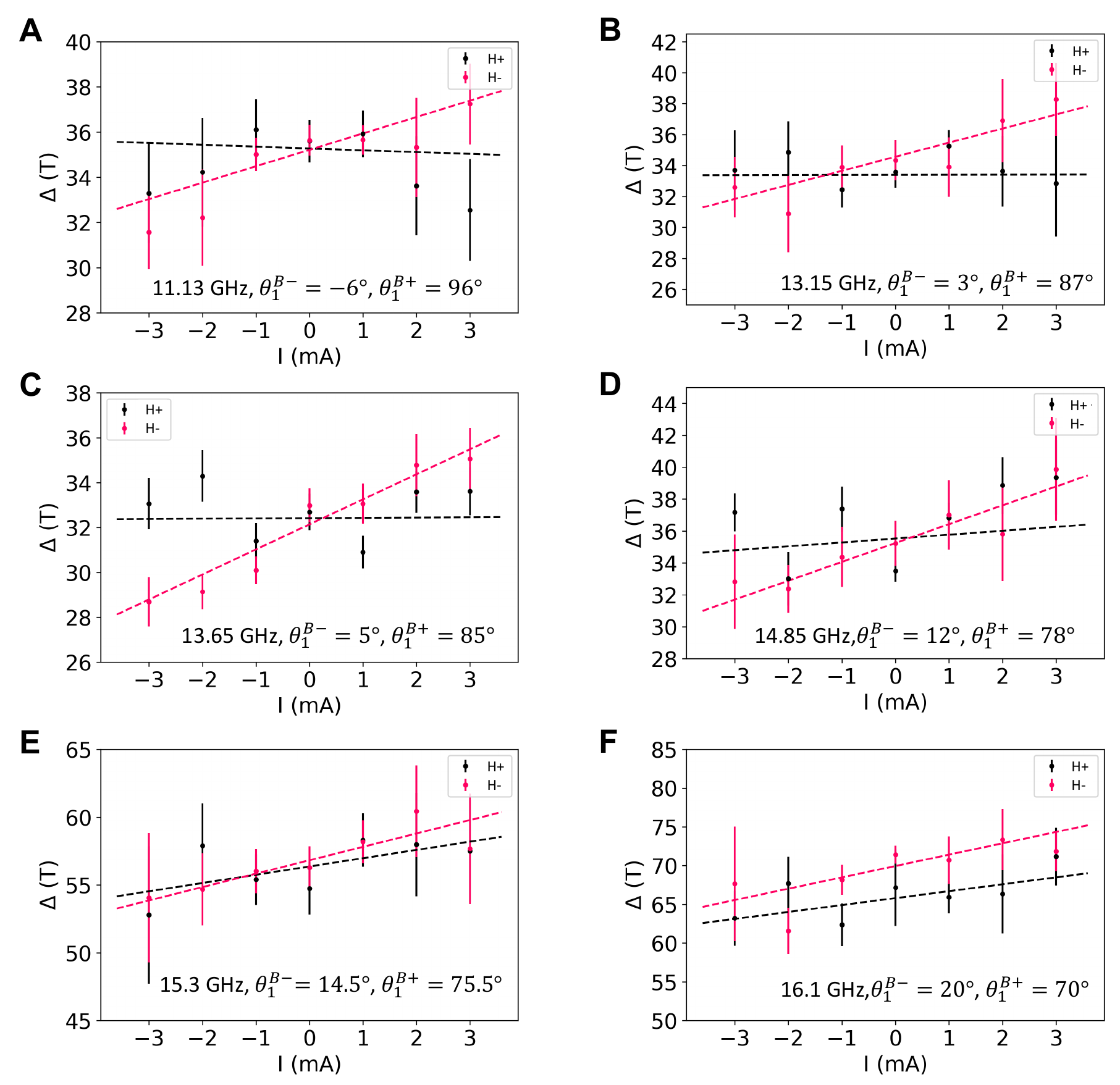}
    \caption{\textbf{Linewidth modulation slopes as a function of dc bias current for device 1 analyzed in the main text.} (\textbf{A}-\textbf{F}) Linear fits to the modulated resonant linewidth at different resonant frequencies, with black and red dashed lines showing fits for positive and negative fields respectively. The fitted slopes have a cos$\theta^{H\pm}_{1}$ dependence (main text Fig.\ 4C). Uncertainties in the linewidth $\Delta$ were determined from the standard deviation errors in the Lorentzian fits to the resonances.}
    \label{FigureS4} 
\end{figure}

\begin{figure}[htpb]
    \centering
    \includegraphics[width=0.65\textwidth]{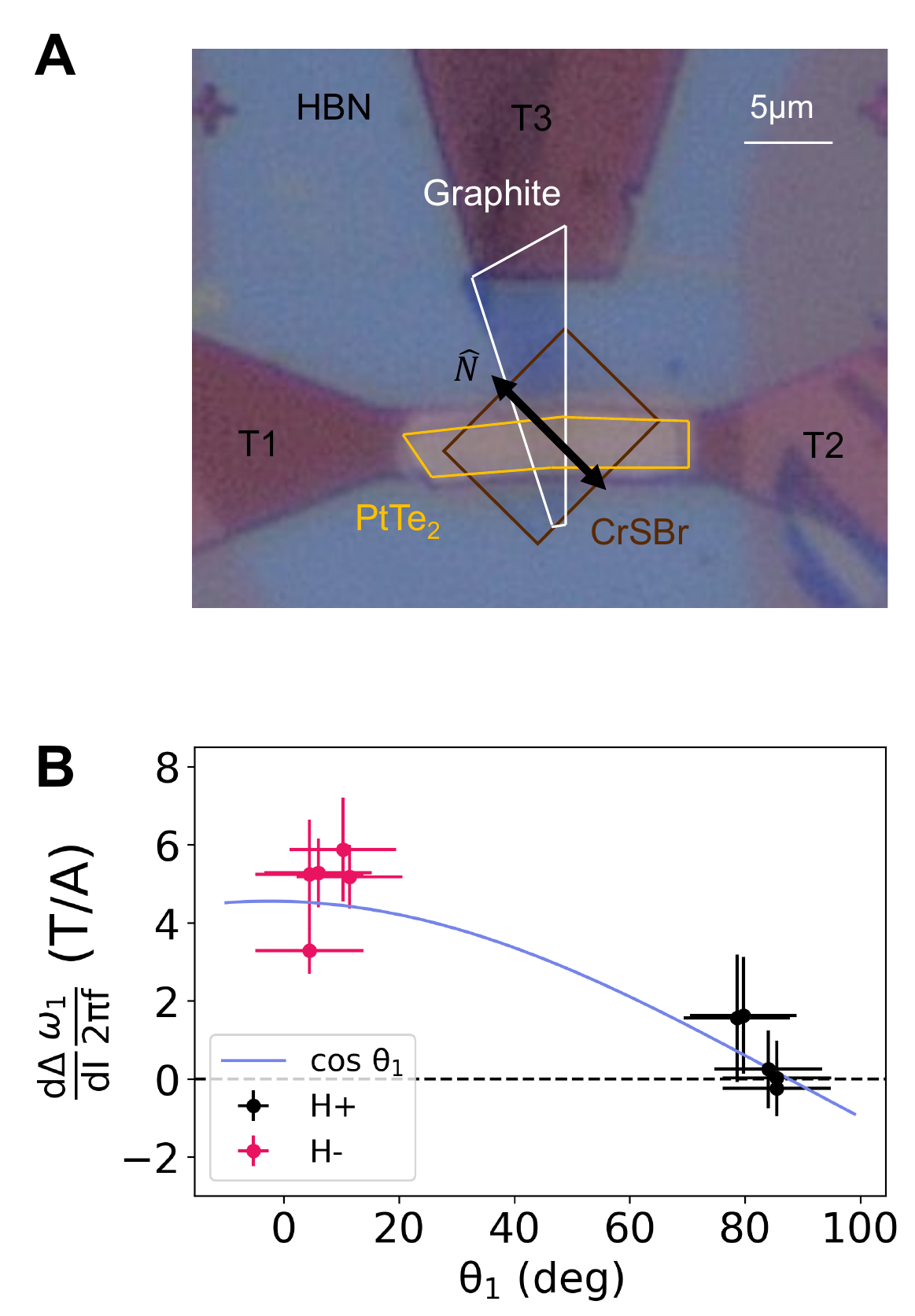}
    \caption{\textbf{Image and data for device 2
 (PtTe2(61.1 nm)/CrSBr(2L))} (\textbf{A}) Image of the hBN/graphite/CrSBr/PtTe$_2$ device 2. (\textbf{B}) Angular dependence of dc-current-modulated linewidths for device 2. The dc-bias-current-induced resonant linewidth modulation follows an approximate $\cos{\theta_1}$ dependence similar to device 1 in the main text. Uncertainties in the dc-bias modulated linewidth $\frac{d\Delta}{dI}$ were determined from the standard deviation errors of linear fits to $\Delta$ v.s. I. Uncertainties in $\theta_1$ were determined from the width in magnetic field of the corresponding resonance.}
    \label{FigureS5} 
\end{figure}

\begin{figure}[htpb]
    \centering
    \includegraphics[width=0.95\textwidth]{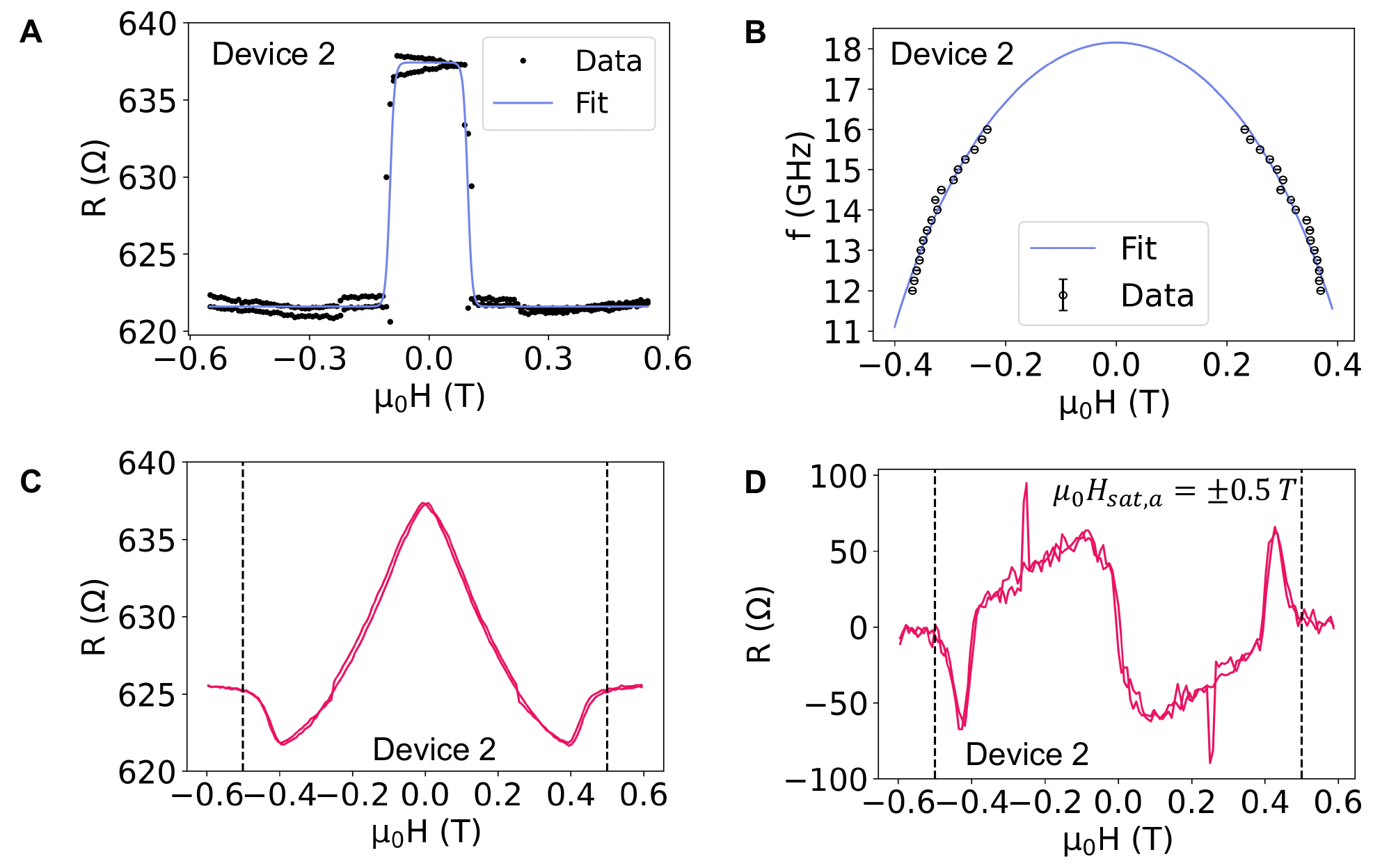}
    \caption{\textbf{Anisotropy and exchange parameters of device 2.} (\textbf{A}, \textbf{B}) Simultaneous fits to the magnetoresistance and resonance spectrum of device 2 at 85 K using S.I.\ Eqs.\ S1 and S2, from which we extract the magnetic anisotropy and exchange parameters $\rm{\mu_0 H_a}$ = 0.31(2) T, $\rm{\mu_0 H_c}$ = 0.83(6) T, and $\rm{\mu_0 H_E}$ = 0.0975(6) T. (\textbf{C}, \textbf{D}) Magnetoresistance and first derivative of the magnetoresistance as a function of external magnetic field along the $\hat{a}$ axis. Dashed lines indicate the estimated saturation magnetic field of $\mu_0H_{sat, a}$ $\approx$ 0.5 T.}
    \label{FigureS6}
\end{figure}

\begin{figure}[htpb]
    \centering
    \includegraphics[width=1.0\textwidth]{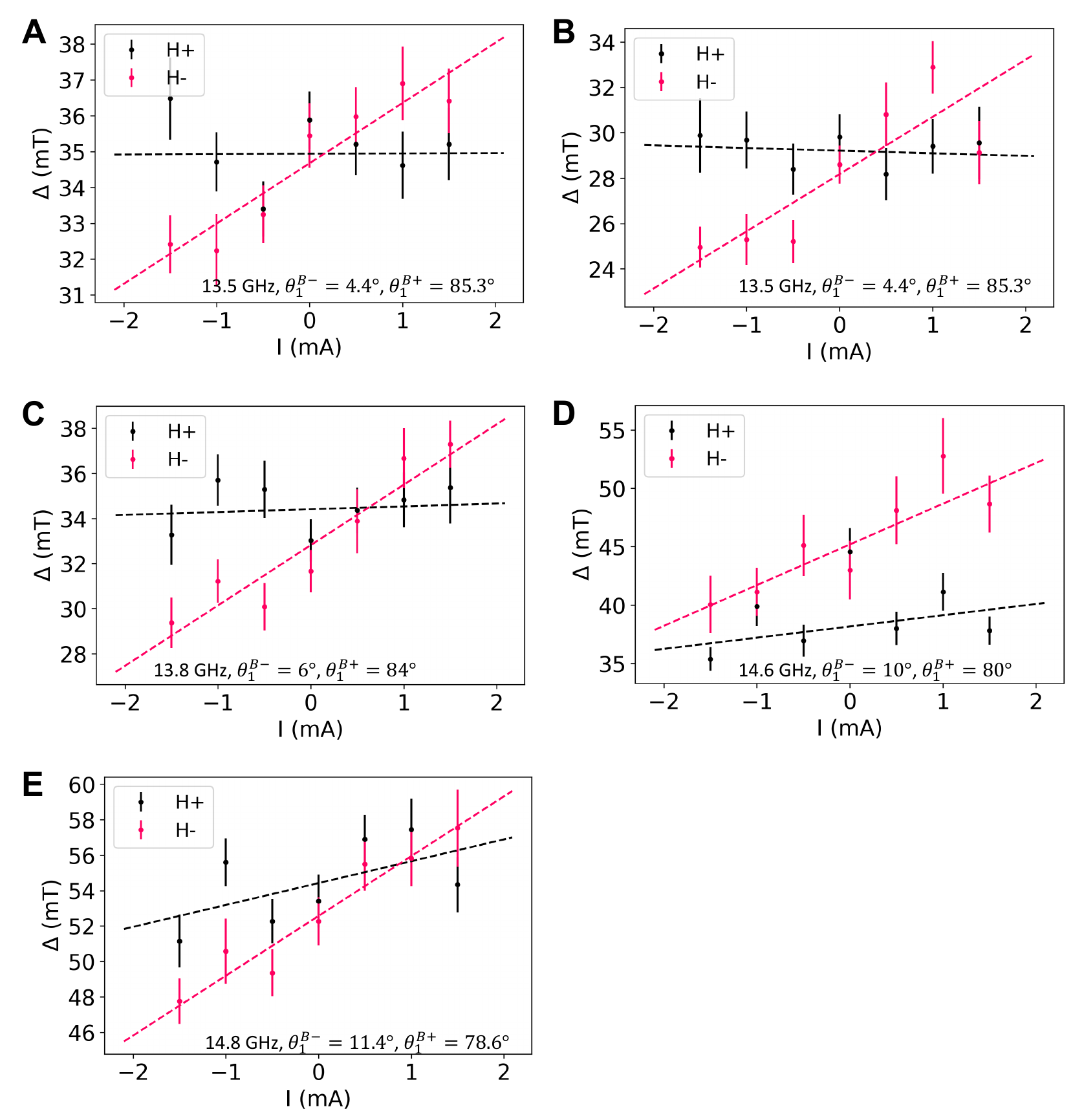}
    \caption{\textbf{Linewidth modulation slopes as a function of dc bias current for device 2 (PtTe$_2$(61.1 nm)/CrSBr(2L)) at 85 K.} (\textbf{A}-\textbf{E}) Linear fits to the modulated resonant linewidth at different resonant frequencies, with black and red dashed lines showing fits for positive and negative fields respectively. The fitted slopes have a cos$\theta^{H\pm}_{1}$ dependence consistent with device 1.}
    \label{FigureS7} 
\end{figure}

\noindent In our measurements of the tunneling resistance as a function of field parallel to the easy axis of CrSBr, we observe an abrupt spin-flip transition from an anti-parallel state to a parallel state. Here, the spin-flip field is equal to the exchange parameter: $\rm{\mu_0 H_{sf} = \mu_0 H_{E}}$\cite{Cham2022}, and the resistance can be expressed as a step function:
\begin{align}
\begin{split}
    \rm{R} &= \rm{R_0 + \Delta R/(1 + e^{-\beta(|\mu_0 H| - |\mu_0 H_E|)})}
\end{split}
\label{eqn:stepfunc}
\end{align}
\noindent where R$_0$ is the resistance at zero field, $\rm{\Delta R}$ is the change in resistance, $\beta$ is a parameter characterizing the slope of the transition, $\rm{\mu_0 H}$ is the external field along the easy axis of CrSBr, and $\rm{\mu_0 H_E}$ is the exchange parameter. In addition, the dependence of frequency versus applied magnetic field for the out-of-phase optical mode detected in the ST-AFMR spectrum (for field applied along the intermediate axis) is given by\cite{Cham2022}:
\begin{align}
\begin{split}
    \rm{\omega_0} &= \rm{\mu_0 \gamma \sqrt{\frac{((H_a + 2H_E)^2 - H^2)H_c}{H_a + 2H_E}}}
\end{split}
\label{eqn:op_mode}
\end{align}
\noindent From simultaneous fits to the tunneling magnetoresistance and the ST-AFMR spectrum (Figs.\ \ref{FigureS2}A,B), we extract the magnetic anisotropy and exchange parameters $\rm{\mu_0 H_a}$ = 0.33(6) T, $\rm{\mu_0 H_c}$ = 0.77(2) T, and $\rm{\mu_0 H_E}$ = 0.096(1) T for device 1 at 85 K. 

\vspace{0.5cm}
\noindent We also verified our value for $\rm{\mu_0 H_a}$ based on the magnetoresistance field sweep along the $\hat{a}$ axis. We identify the saturating field from where the resistance saturates, corresponding to positions where $\rm{dR/dB}$ approaches 0 (Figs.\ \ref{FigureS2}C,D). This gives $\rm{\mu_0H_{sat,a}}$ $\approx$ 0.49 T, and using the relation that the saturation field should equal $\rm{\mu_0H_{sat,a}} = \rm{2\mu_0H_E} + \rm{\mu_0H_a}$, along with the value we determined for the exchange field $\rm{\mu_0H_{E}} = 0.096$ T, we arrive at a value of $\rm{\mu_0H_{a}}$ $\approx$ 0.3 T.  This is close to the value obtained from fitting the magnetic field dispersion. A large positive magnetoresistance component from the PtTe$_2$ for field sweeps along the $\hat{c}$-axis (Fig.\ \ref{FigureS2}E) precludes a similar analysis for $\rm{\mu_0H_{c}}$. Our exchange and anisotropy values $\rm{\mu_0 H_E}$, $\rm{\mu_0 H_a}$ and $\rm{\mu_0 H_c}$ are consistent with other magnetoresistance measurements performed along the $\hat{a}$, $\hat{b}$, and $\hat{c}$ axes in bilayer CrSBr \cite{boix2022probing}.

\vspace{0.25cm}
\subsection{Damping parameter of C\MakeLowercase{r}SB\MakeLowercase{r}}
\noindent We determine the damping parameter of CrSBr from the frequency dependence of the resonance linewidth at zero current bias:
\begin{align}
    \begin{split}
        \Delta  = \alpha \frac{\omega \omega_2}{\gamma\omega_1} + \Delta_0
    \end{split}
\end{align} (see Eq.\ \ref{eqn:AFMDelta}) where $\Delta_0$ allows for a contribution to the linewidth from inhomogeneous broadening. A linear fit to the linewidth $\Delta$ as a function of $\frac{\omega \omega_2}{\gamma \omega_1}$ (Fig.\ \ref{FigureS3}C) gives a damping parameter of $\alpha = 0.066(2)$ and an inhomogeneous broadening parameter of $\Delta_0 = 1.9(1.3)$ mT.

\vspace{0.25cm}
\subsection{Landau-Lifshitz-Gilbert-Slonczewski equations for an antiferromagnet with triaxial anisotropy}
\noindent Following the framework established for ferromagnets\cite{karimeddiny2021resolving, Mittelstaedt2022}, we use a similar formalism to derive analytical solutions for the linewidth modulation of an antiferromagnet in the presence of spin-orbit torques generated by a dc spin current. To account for the additional exchange and anisotropy parameters present in an antiferromagnetic system, we adapt the formalism first introduced in \cite{Macneill2019} and later generalized for triaxial crystalline anisotropy in \cite{Cham2022}. We start with the LLGS equation with unit magnetizations $\bm{m}_{1,2}$ for spin sublattices 1 and 2 and a damping-like torque acting only on sublattice 1:
\begin{align}
    \begin{split}
    \dot{\bm{m}}_{1} &= -\mu_0\gamma \bm{m}_{1} \times \bm{H}_{eff, 1} + \alpha \bm{m}_{1} \times \dot{\bm{m}}_{1} + \tau_{SH} \bm{m}_{1} \times (\hat{\sigma} \times \bm{m}_{1})\\
    \dot{\bm{m}}_{2} &= -\mu_0\gamma \bm{m}_{2} \times \bm{H}_{eff, 2} + \alpha \bm{m}_{2} \times \dot{\bm{m}}_{2}
    \end{split}
    \label{eqn:coupledLLGS}
\end{align} where $\mu_0$ is the vacuum permeability, $\gamma$ is the electron gyromagnetic ratio, $\alpha$ is the Gilbert damping parameter $\tau_{SH}$ is the spin-Hall damping-like torque parameter, and $\hat{\sigma}$ is the spin polarization corresponding to the spin-orbit torque. $\bm{H}_{eff, 1, 2}$ are the effective fields accounting for the applied external magnetic field  H, in-plane and out-of-plane anisotropies H$\rm{_a}$ and H$\rm{_c}$ and the interlayer antiferromagnetic exchange H$_E$. In general, the dynamics are quite complicated. However, we can consider the special case when the external field is applied parallel to the in-plane intermediate anisotropy axis $\hat{a}$ where we can use the symmetry of the spin-flopped sublattices about the intermediate axis. We assume an oscillatory expression for the magnetization of the form $\bm{m}_{1(2)}(t) = \bm{m}_{1(2)}^{eq} + \delta\bm{m}_{1(2)}e^{i\omega t}$. Including terms to first order in precession amplitudes, the coupled LLGS equations for the antiferromagnetic sublattices $\bm{m}_{1(2)}$ are:
\begin{align}
\begin{split}
    i\omega\delta \bm{m}_1 &= -\mu_0\gamma \bm{m}_1^{eq} \times (H_{eq}\delta \bm{m}_1 + H_E\delta \bm{m}_2 + H_c(\delta \bm{m}_1 \cdot {\hat{c}}){\hat{c}} + H_a(\delta \bm{m}_1 \cdot {\hat{a}}){\hat{a}}) \\
    &+ i\omega\alpha \bm{m}_1^{eq} \times \delta \bm{m}_1 + \tau_{SH} \bm{m}_1^{eq} \times (\hat{\sigma} \times \delta\bm{m}_1)\\
    i\omega\delta \bm{m}_2 &= -\mu_0\gamma \bm{m}_2^{eq} \times (H_{eq}\delta \bm{m}_2 + H_E\delta \bm{m}_1 + H_c(\delta \bm{m}_2 \cdot {\hat{c}}){\hat{c}} + H_a(\delta \bm{m}_2 \cdot {\hat{a}}){\hat{a}}) \\
    &+ i\omega\alpha \bm{m}_2^{eq} \times \delta \bm{m}_2.
\end{split}
\label{eqn:linearcoupledLLGS}
\end{align}
\noindent Here, $\bm{m}_{1(2)}^{eq}$, $\delta\bm{m}_{1(2)}^{eq}$ are the equilibrium and displacement vectors of each sublattice, respectively.

\vspace{0.5cm}
\noindent To further simplify the expression, we note that the effective field magnitude $H_{eq}$ acting on a given spin sublattice is simply the magnitude of total effective magnetic field at the equilibrium condition, and given the equilibrium condition this field is parallel to the sublattice moment on which it acts: $H_{eq}\bm{m}_{1(2)}^{eq}=H_\perp \hat{a} - H_E\bm{m}_{2(1)} - H_a(\bm{m}_{1(2)} \cdot \hat{a})\hat{a}$. By balancing the torques at equilibrium and using a few steps of trigonometry, $H_{eq} = |H_\perp{\hat{a}} - H_E\bm{m}_2 - H_a(\bm{m}_1\cdot {\hat{a}}){\hat{a}}| = H_E$. Furthermore, because the spin-flop configuration is 2-fold symmetric about $\hat{a}$, C$_{2a}\bm{\mhat}_1^{eq}$ = $\bm{\mhat}_2^{eq}$, where C$_{2a}$ is an operator for a 2-fold rotation about the $\hat{a}$ axis, we can decouple Eqs.\ \ref{eqn:linearcoupledLLGS} by defining two orthogonal modes that are linear combinations of the sublattice displacement vectors:  $\delta$$\bm{m}_{\pm}$ = $\delta\bm{m}_1$ $\pm$ C$_{2a}$$\delta\bm{m}_2$, and we get the LLG equations rewritten in the basis of $\delta$$\bm{m}_{\pm}$:
\begin{align}
\begin{split}
    i\omega\delta \bm{m}_{\pm} &= -\mu_0\gamma \bm{m}_1^{eq} \times (H_{e}(\delta \bm{m}_{\pm} \pm C_{2a}\delta \bm{m}_{\pm}) + H_c(\delta \bm{m}_{\pm} \cdot {\hat{c}}){\hat{c}} + H_a(\delta \bm{m}_{\pm} \cdot {\hat{a}}){\hat{a}})\\
    &+ i\omega\alpha \bm{m}_1^{eq} \times \delta \bm{m}_{\pm}\\
    &=  i\omega \left( \delta \bm{m}^{eff}_{\pm} + \delta \bm{m}^{\alpha}_{\pm} \right).
\end{split}
\label{eqn:decoupledLLGS}
\end{align} Here $\delta \bm{m}^{eff}_{\pm}$ and $\delta \bm{m}^{\alpha}_{\pm}$ are the torques arising from the effective field and Gilbert damping respectively.  In the absence of the spin-orbit torque, the in-phase ($\delta\bm{m}_{-}$) and out-of-phase ($\delta\bm{m}_{+}$) modes are fully decoupled.\cite{Macneill2019, Cham2022} The spin-orbit torque will excite both modes at the same time, but we make the approximation that this torque is sufficiently weak so that if the driving frequency is close to one of the antiferromagnetic resonance modes but not the other then only one mode will be significantly excited, and hence the problem will remain decoupled to a good approximation.   We will focus on the effect of torques on $\delta\bm{m}_{+}$, the out-of-phase mode that our tunneling measurements detect. Rewriting the spin-orbit torque in Eq.\ \ref{eqn:coupledLLGS} in terms of $\delta\bm{m}_{+}$ and $\delta\bm{m}_{-}$ we get
%as a sum and difference of out-of-phase and in-phase torques and using the relation $\delta\bm{m}_1 = \rm{C_{2a}}$$\delta\bm{m}_2$ for the out-of-phase mode, we get:
\begin{align}
    \begin{split}
    \bm{\tau}_{SH} &= \tau_{SH} \bm{m}_{1}^{eq} \times (\hat{\sigma} \times \delta\bm{m}_{1})\\
    &= \tau_{SH} \bm{m}_{1}^{eq} \times \left(\hat{\sigma} \times \left(\frac{\delta\bm{m}_{+}}{2} + \frac{\delta\bm{m}_{-}}{2}\right)\right).
    \end{split}
\end{align}
If we consider only the out-of-phase mode, we therefore get a factor of $\frac{1}{2}$ for the effective spin-orbit torque acting on $\delta\bm{m}_{+}$:
\begin{align}
    \begin{split}
    \tau_{SH} \bm{m}_1^{eq} \times (\hat{\sigma} \times \delta \bm{m}_{1}) &\rightarrow \frac{\tau_{SH}}{2} \bm{m}_{1}^{eq} \times (\hat{\sigma} \times \delta\bm{m}_{+}).
    \end{split}
\end{align}

\noindent Before proceeding to solve the equations of motion, we can try to get an intuition for the effects of the additional Gilbert damping and damping-like spin-orbit torque terms. We note that the form of the decoupled equations are very similar to those for a single ferromagnetic layer with equilibrium position $\bm{m}_{1}^{eq}$ and deviations from that position $\delta\bm{m}_{\pm}$. Namely, the Gilbert damping torque acts to damp out $\delta\bm{m}_{\pm}$ to zero, whilst the damping-like spin-orbit torque acts either to enhance or reduce this damping. Expanding Eq.\ \ref{eqn:decoupledLLGS}:
\begin{align}
\begin{split}
     i\omega \delta \bm{m}^{eff}_{\pm}
     &= -\mu_0\gamma
    \begin{pmatrix}
       \cos{\chi}\\
       \sin{\chi}\\
       0
    \end{pmatrix} \times \left[
    H_E\left(\begin{pmatrix}
       \delta m_{\pm, x}\\
       \delta m_{\pm, y}\\
       \delta m_{\pm, z}
    \end{pmatrix} \pm 
    \begin{pmatrix}
       -\delta m_{\pm, x}\\
       \delta m_{\pm, y}\\
       -\delta m_{\pm, z}
    \end{pmatrix}\right) + 
    \begin{pmatrix}
       0\\
       H_a\delta m_{\pm, y}\\
       H_c\delta m_{\pm, z} 
    \end{pmatrix}\right]\\
    &= \begin{cases}
        \begin{pmatrix}
       0\\
       0\\
       2H_E\delta m_{+,y}\cos{\chi} 
    \end{pmatrix} + 
    \begin{pmatrix}
       H_c \delta m_{+,z} \sin{\chi}\\
       -H_c \delta m_{+,z} \cos{\chi}\\
       H_a\delta m_{+,y} \cos{\chi} 
    \end{pmatrix} & \text{out-of-phase}\\
    \begin{pmatrix}
       2H_E \delta m_{-,z}\sin{\chi}\\
       -2H_E \delta m_{-,z}\cos{\chi}\\
       -2H_E \delta m_{-,x}\sin{\chi} 
    \end{pmatrix}  + 
    \begin{pmatrix}
       H_c \delta m_{-,z} \sin{\chi}\\
       -H_c \delta m_{-,z} \cos{\chi}\\
       H_a \delta m_{-,y} \cos{\chi} 
    \end{pmatrix} & \text{in-phase}
    \end{cases}
\end{split}
\end{align}

\noindent Taking into consideration that $\delta\bm{m}_+$ and $\delta\bm{m}_-$ are always perpendicular to $\bm{m}_{1}$, i.e. the length of $\bm{m}_{1}$ is unchanged, we can simplify the problem further by expressing the equation in the spherical coordinate basis (r, $\phi$, $\theta$), following the method described in \cite{Macneill2019}. We use the equilibrium position $\sin \chi = H_\perp/(2H_E + H_{a})$ calculated by minimizing the free energy, and adopt the basis ($\delta\bm{m}_{+, \phi}$, $\delta\bm{m}_{+, \theta}$, $\delta\bm{m}_{-, \phi}$, $\delta\bm{m}_{-, \theta}$). 
\begin{align}
\begin{split}
     i\omega \delta \bm{m}^{eff}_{\pm}
     = \begin{cases}
        \begin{pmatrix}
       0\\
       2H_E\delta m_{+,\phi}\cos^2{\chi} 
    \end{pmatrix} + 
    \begin{pmatrix}
       -H_c \delta m_{-,\theta}\\
       H_a \delta m_{-,\phi}\cos^2{\chi} 
    \end{pmatrix}
    & \text{out-of-phase}\\
    \begin{pmatrix}
       -H_E \delta m_{-,\theta}\\
       H_E \delta m_{-,\phi}\sin^2{\chi}
    \end{pmatrix} + 
    \begin{pmatrix}
       -H_c \delta m_{+,\theta}\\
       H_a \delta m_{+,\phi}\cos^2{\chi} 
    \end{pmatrix} & \text{in-phase}
    \end{cases}
\end{split}
\label{eqn:decoupledLLGSeff}
\end{align}
In block diagonal form, Eq.\ \ref{eqn:decoupledLLGSeff} can be written as
\vspace{0.25cm}
\begin{align}
i\omega
\begin{pmatrix}
\delta\bm{m}_{+}\\
\delta\bm{m}_{-}
\end{pmatrix}
=
\begin{pmatrix}
   A & 0 \\
   0 & B 
   \end{pmatrix} 
\begin{pmatrix}
\delta\bm{m}_{+}\\
\delta\bm{m}_{-}
\end{pmatrix}
\label{eqn:blockdiag}
\end{align} where
\begin{align}
\bm{A} &= 
\mu_0\gamma \begin{pmatrix}
 0 & H_c \\
 -(2H_E + H_a)(1 - (H/(2H_E + H_a))^2) & 0 \\
\end{pmatrix}
\end{align} 

\begin{align}
\bm{B} &= 
\mu_0\gamma \begin{pmatrix}
 0 & H_E+H_c \\
 -(H_a - (H_a - H_E)(H/(2H_E + H_a))^2) & 0 \\
\end{pmatrix}.
\end{align} 

\noindent With these simplified matrices, we can include the Gilbert damping and damping-like spin-orbit torque in this basis. Considering only the upper left quadrant of Eq.\ \ref{eqn:blockdiag} for the out-of-phase mode, the LLGS can be written as a 2 $\times$ 2 matrix equation:
\begin{align}
i\omega\begin{pmatrix}
    \delta m_\phi\\
    \delta m_\theta\\ 
\end{pmatrix} = (\bm{A} + \bm{M_{\alpha}} + \bm{M_{\tau_{SH}}}) \times \begin{pmatrix}
    \delta m_\phi\\
    \delta m_\theta\\ 
\end{pmatrix}.
\end{align}
Considering the Gilbert damping term first
\begin{align}
\delta\bm{m}_{\pm}^{\alpha} = i\omega\alpha \begin{pmatrix}
    1\\
    0\\
    0\\
\end{pmatrix} \times \begin{pmatrix}
    0\\
    \delta m_{\phi}\\
    \delta m_{\theta}\\
\end{pmatrix}= i\omega\alpha\begin{pmatrix}
    0 \\ -\delta m_{\theta} \\ \delta m_{\phi}
\end{pmatrix}
\end{align}
\noindent gives
\begin{align}
\bm{M_{\alpha}} = \begin{pmatrix}
    0 & -i\omega\alpha\\
    i\omega\alpha & 0
\end{pmatrix}
\end{align}
\noindent and considering spin polarization within the plane
\begin{align}
    \bm{\tau}_{SH} = \frac{\tau_{SH}}{2} \begin{pmatrix}
    1\\
    0\\
    0\\
    \end{pmatrix} \times \left[\begin{pmatrix}
    \sigma_{r}\\
    \sigma_{\phi}\\
    0\\
    \end{pmatrix} \times \begin{pmatrix}
    0 \\
    \delta m_{\phi} \\
    \delta m_{\theta} \\
    \end{pmatrix}\right] = -\frac{\tau_{SH}}{2} \begin{pmatrix}
    0\\
    \sigma_{r}\delta m_{\phi}\\
    \sigma_{r}\delta m_{\theta}\\
    \end{pmatrix}
\end{align}
\noindent where $\sigma_{r}$ and $\sigma_{\phi}$ are the spin polarization components parallel and perpendicular to $\bm{m}_{1}^{eq}$, to give
\begin{align}
\bm{M}_{\tau_{SH}} = \begin{pmatrix}
    -\frac{\tau_{SH, \parallel}}{2} & 0\\
    0 & -\frac{\tau_{SH, \parallel}}{2}
\end{pmatrix}.
\end{align}

\noindent Putting everything together for the out-of-phase mode in the spherical basis
\begin{align}
    i\omega \begin{pmatrix}
        \delta m_{\phi}\\
        \delta m_{\theta}
    \end{pmatrix} = 
    \begin{pmatrix}
        -i\omega -  \frac{\tau_{SH, \parallel}}{2} & -i\omega\alpha + \mu_0\gamma H_{c}\\
        -\mu_0\gamma(H_{e,a})(1 - (\frac{H}{H_{e,a}})^2) + i\omega\alpha & \ \ \ -i\omega - \frac{\tau_{SH, \parallel}}{2}
    \end{pmatrix}
    \begin{pmatrix}
        \delta m_{\phi}\\
        \delta m_{\theta}
    \end{pmatrix} = 0
\end{align}
\noindent where we have defined $H_{e,a} = (2H_{E} + H_a)$. Solving for the eigenvalues would give us the out-of-phase mode solutions\cite{Cham2022}, modified by the additional Gilbert damping and spin-orbit torque terms. At the same time, we can also deduce the modulation of the linewidth by the damping-like spin-orbit torque. This linewidth is equivalent to the half width at half maximum of the response to a periodic driving force, analogous to a damped harmonic oscillator. We deduce this from the susceptibility matrix of the response to an oscillating torque $\bm{\tau}_{ac}$, given by the inverse of the matrix, $\bm{M}^{-1}$:
\begin{align}
\begin{pmatrix}
    \delta m_{\phi}\\
    \delta m_{\theta}
\end{pmatrix} = 1/det[\bm{M}] \begin{pmatrix}
        -i\omega - \frac{\tau_{SH, \parallel}}{2} & i\omega\alpha - \mu_0\gamma H_{c}\\
        \mu_0\gamma(H_{e,a})(1 - (\frac{H}{H_{e,a}})^2) - i\omega\alpha & \ \ \ -i\omega - \frac{\tau_{SH, \parallel}}{2}
    \end{pmatrix} \begin{pmatrix}
    \tau_{ac, \phi}\\
    \tau_{ac, \theta}
\end{pmatrix}.
\end{align}

\noindent The prefactor of the susceptibility matrix is:
\begin{align}
\begin{split}
    1/det[\bm{M}] &= \frac{1}{\frac{(\mu_0\gamma)^2H_c(H_{e,a}^2 - H^2)}{H_{e,a}} + \frac{\tau_{SH, \parallel}}{2}^2 - (1+\alpha^2)\omega^2 - i\omega[\mu_0\gamma\frac{H_{e,a}(H_c + H_{e,a}) - H^2}{H_{e,a}}\alpha + \tau_{SH, \parallel}]} \\
    &= \frac{1}{\omega_0^2 - (1 + \alpha^2)\omega^2 + \frac{\tau_{SH, \parallel}}{2}^2 - i\omega[\omega_{2}\alpha + \tau_{SH, \parallel}]} \\ 
        &\approx \frac{1}{\omega_0^2 - \omega^2 - i\omega[\omega_{2}\alpha + \tau_{SH, \parallel}]} \\ 
        &= \frac{(\omega_0^2 - \omega^2) + i\omega(\omega_{2}\alpha + \tau_{SH, \parallel})}{(\omega_0^2 - \omega^2)^2 + [\omega(\omega_{2}\alpha + \tau_{SH, \parallel})]^2}
        \label{eqn:AFMRSusceptibility}
    \end{split}
\end{align} where we have defined $\omega_{0}^2 = (\mu_0\gamma)^2\frac{H_c(H_{e,a}^2 - H^2)}{H_{e,a}}$, $\omega_{2} = \mu_0\gamma\frac{H_{e,a}(H_c + H_{e,a}) - H^2}{H_{e,a}}$, and taken damping and torque terms to linear order. Here, we check that $\omega_0$ is simply the resonant frequency as a function of field for the out-of-phase mode, such that the magnetization response is maximized when the driving frequency is equal to the resonant frequency ($\omega^2 - \omega_0^2$ = 0). 

\vspace{0.5cm}
\noindent The susceptibility expression looks a little complicated, but has a similar functional form as that for a ferromagnet\cite{Mittelstaedt2022}. Following a similar treatment, we can rewrite the frequency variables in the form of field variables:
\begin{align}
    \begin{split}
        \omega_0^2(H) &= \omega_0^2(H_0) + (H-H_0)\frac{d\omega_0^2}{dH}\Bigr|_{H = H_0}\\
        &= \omega^2 + (H-H_0) \frac{d}{dH}\Bigr|_{H = H_0} \gamma^2 \mu_0^2 \frac{(H_{e,a}^2 - H^2)H_c}{H_{e,a}}\\
        &= \omega^2 - 2(H-H_0)\gamma^2 \mu_0^2 \frac{H_c}{H_{e,a}}H_0\\
        &= \omega^2 - \gamma \omega_1 \mu_0 (H - H_0)
    \end{split}
\end{align}
\noindent where $\omega_1 = 2 \mu_0 \gamma \frac{H_c}{H_{e,a}}H_0$. Substituting this into Eq.\ \ref{eqn:AFMRSusceptibility}:
\begin{align}
\begin{split}
    1/det[\bm{M}] &= \frac{-\gamma\omega_1\mu_0 (H - H_0) + i\omega(\omega_{2}\alpha + \tau_{SH, \parallel})}{(\gamma\omega_1\mu_0)^2(H - H_0)^2 + [\omega(\omega_{2}\alpha + \tau_{SH, \parallel})]^2}\\
    &= \frac{1}{\gamma\omega_1}\frac{-\mu_0 (H - H_0) + i\frac{\omega}{\gamma\omega_1}(\omega_{2}\alpha + \tau_{SH, \parallel})}{\mu_0^2 (H - H_0)^2 + [\frac{\omega}{\gamma\omega_1}(\omega_{2}\alpha + \tau_{SH, \parallel})]^2}\\
    &= \frac{1}{\gamma\omega_1}\frac{-\mu_0 (H - H_0) + i\Delta}{\mu_0^2 (H - H_0)^2 + \Delta^2}
    \end{split}
\end{align}
\noindent where we have defined the linewidth:
\begin{align}
\Delta = \frac{\omega}{\gamma\omega_1}(\omega_2\alpha + \tau_{SH, \parallel}).
\label{eqn:AFMDelta}
\end{align} 

\noindent The amplitude of the spin-orbit torque scales linearly with the applied dc-bias current I$_{dc}$:\cite{liu2011spin, karimeddiny2021resolving}
\begin{align}
    \begin{split}
        \tau_{SH, \parallel} = \frac{\hbar}{2e}\frac{\gamma  \xi_{SH}  \cos{\phi_{\sigma}}}{M_s W t_m t_{nm}} I_{dc}
    \end{split}
    \label{eqn:torque}
\end{align}
\noindent where $\xi_{SH}$ is the spin-orbit torque efficiency, $\phi_{\sigma}$ is the relative angle between $\bm{m}_1^{eq}$ and the spin polarization $\hat{\sigma}$, M$_s$ is the saturation magnetization, W is the width of the PtTe$_2$ channel, $t_{nm}$ is the thickness of the PtTe$_2$ channel and $t_m$ is the thickness of a CrSBr monolayer. Substituting this into the expression for the $\Delta$, we get:
\begin{align}
    \begin{split}
        \frac{d\Delta}{dI_{dc}} &= \frac{\omega}{\gamma\omega_1}\frac{\hbar}{2e}\frac{\gamma\xi_{SH}\cos{\phi_\sigma}}{M_s W t_{m} t_{nm}}\\ &= \left(\frac{\hbar\xi_{SH}}{2 e M_s W t_{m} t_{nm}}\right)\frac{\omega}{\omega_1}\cos{\phi_{\sigma}}.
    \end{split}
\end{align}
\noindent Rearranging, the spin-orbit torque efficiency is given by:
\begin{align}
    \begin{split}
        \xi_{SH} &= \frac{2 e M_s W t_{m} t_{nm}}{\hbar}\frac{\omega_1}{\omega \cos_{\phi_\sigma}} \frac{d\Delta}{dI_{dc}}.
    \end{split}
    \label{eqn:AFMRDcbiasXi}
\end{align}

\vspace{0.25cm}
\subsection{Regarding fits to the resonance lineshapes}
\noindent The resonance lineshapes can computed within the same LLGS formalism described in the previous section by including oscillatory torques, analogous to  conventional ST-FMR studies of ferromagnetic films\cite{liu2011spin}. The resonance peaks in the mixing voltage take the form of a sum of symmetric and antisymmetric Lorentzians: 
\begin{align}
    \rm{V^{mix}_{dc}} = \rm{c_S\frac{\Delta^2}{\Delta^2 + (\mu_0H - \mu_0H_0)^2} + c_A\frac{\mu_0H\Delta}{\Delta^2 + (\mu_0H - \mu_0H_0)^2}}
\label{eqn:Lorentzians}
\end{align}
 where $\Delta$ is the half width at half maximum of the Lorentzians, H$_0$ is the resonant field, and c$_S$ and c$_A$ are fitting weights. In conventional two-terminal ST-FMR, the symmetric and antisymmetric components are associated with the damping-like and field-like torque components, respectively\cite{liu2011spin}. However, in three-terminal ST-FMR measurements this simple separation is no longer possible, as parasitic capacitances can give rise to a phase difference between the microwave currents in the PtTe$_2$ channel and within the tunnel junction which shifts the relative amplitudes of the symmetric and antisymmetric contributions\cite{xue2012resonance}.

\vspace{0.25cm}
\subsection{LLGS numerical simulations}
\noindent To better understand the asymmetry in the linewidth modulation effects for positive and negative fields in our measurements, we perform macrospin LLGS numerical calculations. We again consider the case where all of the torques are preferentially acting on sublattice 1:
\begin{align}
\begin{split}
    \dot{\bm{m}}_{1} &= -\mu_0\gamma \bm{m}_{1} \times \bm{H}_{eff,1} + \alpha \bm{m}_{1} \times \dot{\bm{m}}_{1} + \tau_{SH} \bm{m}_{1} \times (\hat{\sigma} \times \bm{m}_{1}).\\
    \dot{\bm{m}}_{2} &= -\mu_0\gamma \bm{m}_{2} \times \bm{H}_{eff,2} + \alpha \bm{m}_{2} \times \dot{\bm{m}}_{2}
\end{split}
\end{align}
\noindent with the external field along the intermediate $\hat{a}$ axis:
\begin{align}
H_{eff, (1,2)} = H_{\perp}{\hat{a}} + H_E\delta \bm{m}_{(2,1)} + H_c(\delta \bm{m}_{(1,2)} \cdot {\hat{c}}){\hat{c}} + H_a(\delta \bm{m}_{(1,2)} \cdot {\hat{a}}){\hat{a}}).
\end{align}

\noindent We numerically integrate these equations to calculate $\bm{m}_{1}$ and $\bm{m}_{2}$ as a function of time, while sweeping the external field. We then calculate the time dependent angle between the two sublattices $\phi$, from which we can obtain the mixing voltage:
\begin{align}
\begin{split}
V_{mix}(t) &= I_{RF} \sin{(\omega t)} \times [R_0 + \Delta R\ 
\cos{(\phi (t))}]\\
        &= V^{dc}_{mix} + O(2\omega t).
\end{split}
\end{align}
\noindent The dc component of the mixing voltage can be compared to the rectified voltage in our ST-AFMR measurements. As seen in Fig.\ \ref{FigureS9}A the amplitude of oscillations in $\cos{\phi}$ is maximum when the antiferromagnet is on resonance. This can be detected as a peak in $V^{dc}_{mix}$ (Fig.\ \ref{FigureS9}B) on resonance. We repeat these time-dependent simulations at various external magnetic fields and positive and negative dc-bias currents to understand how the current-induced spin-orbit torque modifies the linewidth. We consider the scenario where negative fields cant sublattice 1 parallel to the spin polarization $\sigma$ while positive fields cant sublattice 1 perpendicular to $\sigma$ (Fig.\ \ref{FigureS9}C and D). As shown in Fig.\ \ref{FigureS9}E and F, the simulated resonance peaks show an increased(decreased) linewidth for positive(negative) dc bias current for negative fields. For positive fields, we see minimal changes in the linewidth for positive and negative dc bias currents. These simulation results are in agreement with our measurements, supporting our conclusion that spin-orbit torques from the PtTe$_2$ layer in our heterostructure preferentially affect just one of the two CrSBr antiferromagnetic sublattices. The amplitude of the simulated linewidth modulation (main text Fig.\ 4D) agrees with the analytical expression derived in the previous section (Eq.\ \ref{eqn:AFMDelta}).

\vspace{0.25cm}
\subsection{Effect of reversing the N\'eel vector on linewidth modulation}
\begin{figure}[htpb]
    \centering
    \includegraphics[width=1.0\textwidth]{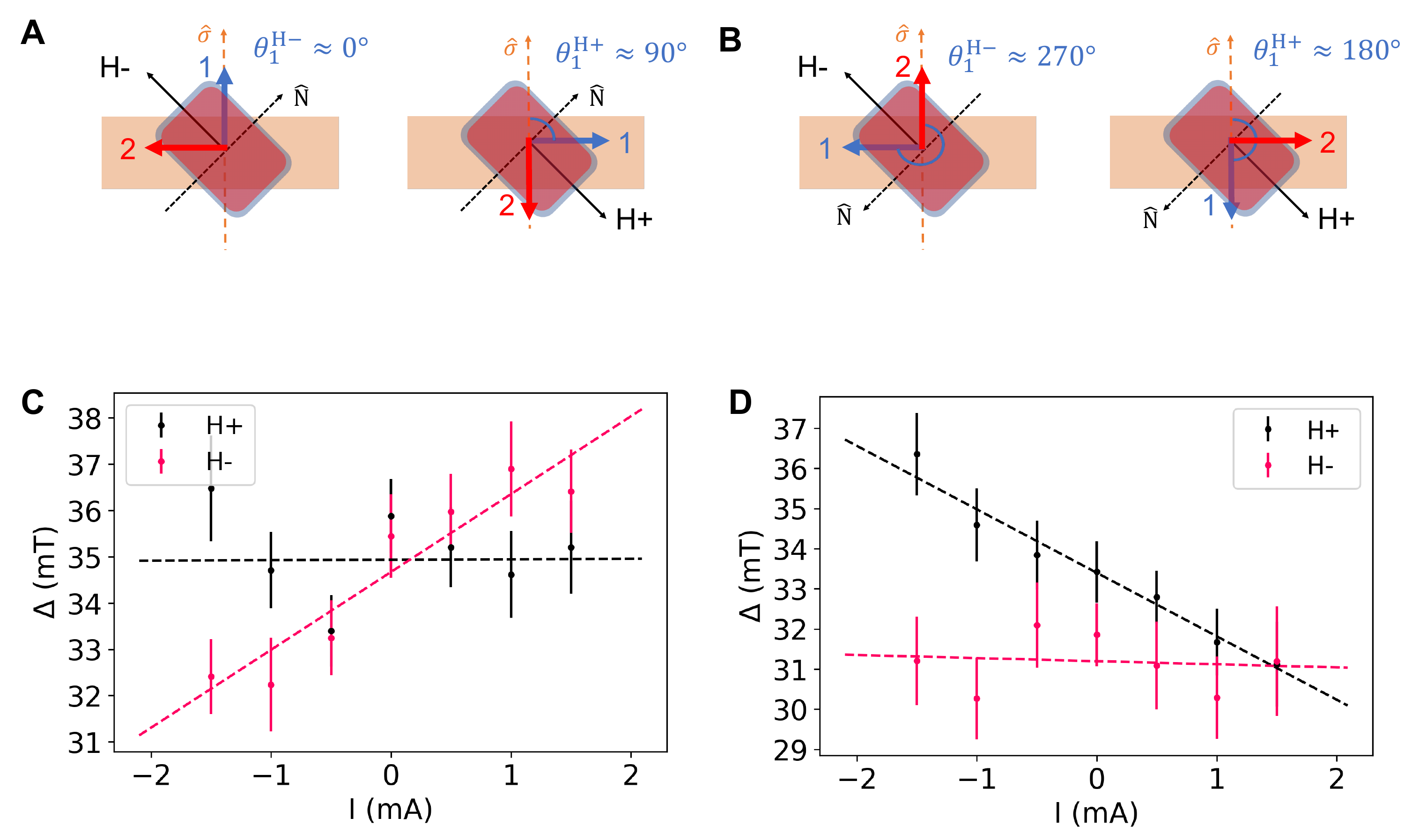}
    \caption{\textbf{Comparison of linewidth modulation for the spin sublattices initialized in opposite states for device 2, analogous to main text Figs.\ 3E,F for device 1.} (\textbf{A}, \textbf{B}) Illustration of the antiferromagnetic sublattices 1 and 2 canted by positive (H+) and negative (H-) fields when sublattice 1 is initialized in the (\textbf{A}) upper right quadrant and (\textbf{B}) lower left quadrant. Black and orange dashed arrows indicate the initial orientation of the N\'eel vector $\hat{N}$ and spin polarization $\sigma$ respectively. (\textbf{C}) Measured resonance linewidth as a function of dc bias current, corresponding to the situation depicted in (\textbf{A}). (\textbf{D}) Resonance linewidth as a function of dc bias current when sublattice 1 is initialized in the opposite direction, corresponding to the situation in (\textbf{B}). The magnitude of the slopes for H+ and H- in Fig.\ \ref{FigureS8}C agree with the slope magnitudes for H- and H+ respectively in Fig.\ \ref{FigureS8}D within uncertainties.}
    \label{FigureS8}
\end{figure}

\begin{figure}[htpb]
    \centering
    \includegraphics[width=0.85\textwidth]{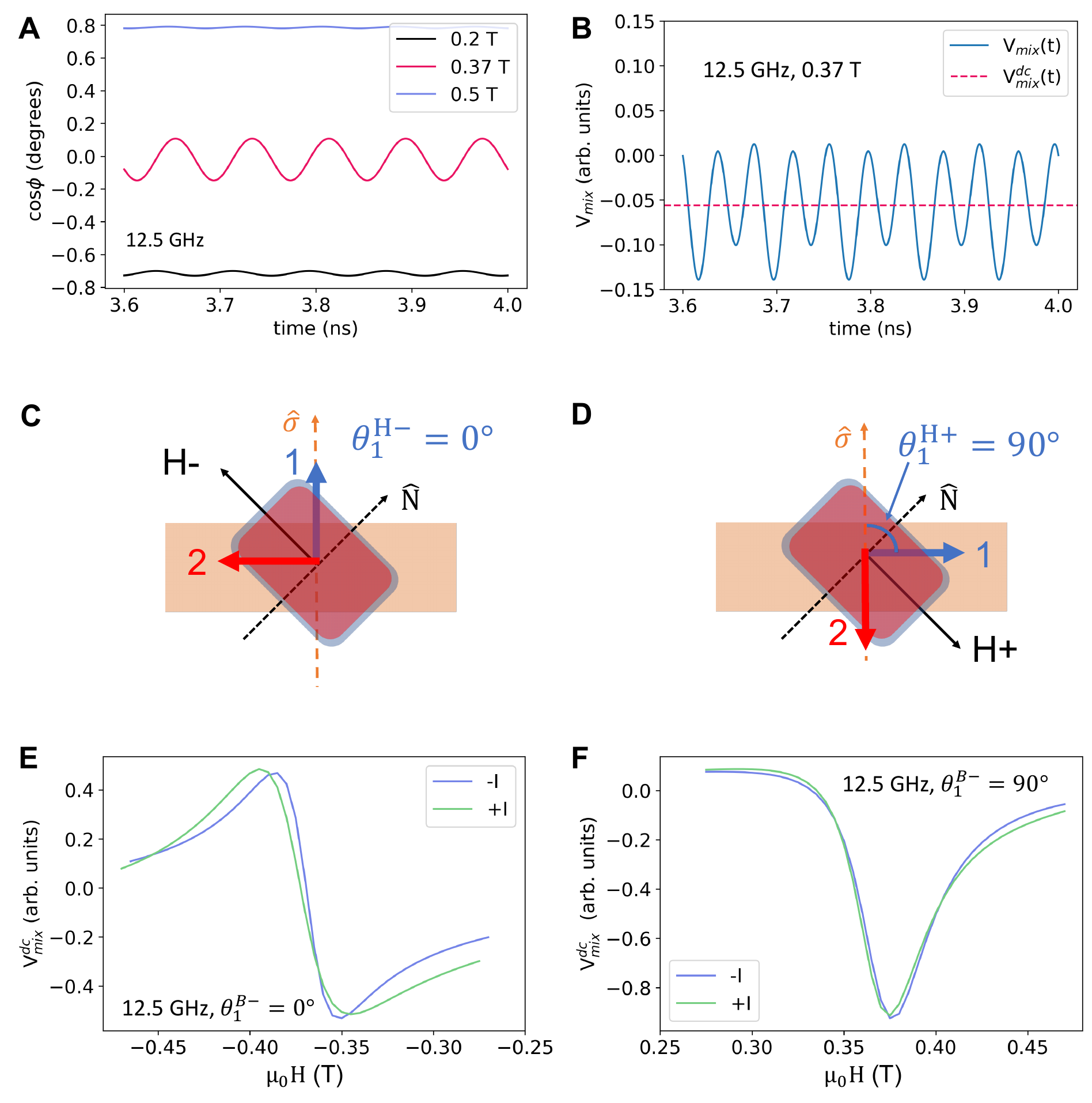}
    \caption{\textbf{Simulated antiferromagnetic dynamics from numerical integration of the coupled LLGS equations.} Simulation parameters from fits to experimental data, $\rm{\mu_0H_a}$ = 0.33 T, $\rm{\mu_0H_E}$ = 0.096 T, $\rm{\mu_0H_c}$ = 0.77 T, were used for the calculations. Damping-like spin-orbit torques were exerted only on sublattice 1. (\textbf{A}) Time evolution of cos($\phi$) for 3 different field values. (\textbf{B}) Mixing voltage on resonance, where the rectified dc component of the mixing voltage is shown by the red dashed line. (\textbf{C}, \textbf{D}) Schematic of the simulated system, with external fields perpendicular to the easy axis of antiferromagnetic sublattices 1 and 2 in the (\textbf{C}) negative and (\textbf{D}) positive directions. (\textbf{E}, \textbf{F}) $V^{dc}_{mix}$ vs.\ external field at negative (-I) and positive (+I) dc bias currents.}
    \label{FigureS9} 
\end{figure}

\noindent We further verify that the linewidth modulation effects observed arise from spin-orbit torques by measuring the effects on the same CrSBr/PtTe$_2$ heterostructure (device 1) with the antiferromagnetic sublattices initialized in the opposite configuration. We do this by sweeping the magnetic field along the easy axis of CrSBr in either the positive or negative field, i.e., parallel or antiparallel to  the black dashed N\'eel-vector arrow $\hat{N}$ in Fig.\ \ref{FigureS8} A and B. This initializes the CrSBr sublattices to one of two near-degenerate states with the interfacial CrSBr layer in opposite starting orientations. Experimentally, we then observe that for the different initial conditions the current-induced linewidth modulation has different dependencies on sign of magnetic field subsequently applied along the intermediate anisotropy axis. In the first case (main text Fig.\ 3E) an H- field gives a current dependence with a positive slope, whilst the current dependence is negligible with an H+ field. When the sublattices are initialized in the opposite direction, we see that an H- field now gives a negligible current dependence while the H+ field gives a current dependence with a negative slope (main text Fig.\ 3F). This is as expected for the different orientations of sublattice 1 with respect to $\sigma$ for the two cases.

\vspace{0.5cm}
\noindent We consider the different configurations of the sublattices when canted by the H- and H+ fields in more detail. In the first configuration depicted in Fig.\ \ref{FigureS8}A, the angle $\theta_{1}$ between $\hat{\sigma}$ and sublattice 1 is 0$^{\circ}$ for H- and 90$^{\circ}$ for H+. Given that the spin-orbit torque induced linewidth modulation follows a $\cos\theta_1$ relation, we indeed measure a positive slope for H- and a negative slope for H+. When the sublattices are flipped into the second configuration, the angles become -90$^{\circ}$ for H- and 180$^{\circ}$ for H+ (Fig.\ \ref{FigureS8}B). In agreement with the theory, we now observe negligible current induced slope for H- and a negative slope for H+. Furthermore, the amplitudes of the positive slope of $\rm{d\Delta/dI}$ = 1.1(2) T/A for H- in main text Fig.\ 3E and the negative slope of $\rm{d\Delta/dI}$ = 1.07(17) T/A in main text Fig.\ 3F agree within uncertainties. We observe a similar effect in device 2, with a positive slope of $\rm{d\Delta/dI}$ = 1.7(3) T/A for H- in Fig.\ \ref{FigureS8}C and a negative slope of $\rm{d\Delta/dI}$ = 1.6(1) T/A in Fig.\ \ref{FigureS8}D. These observations further support our conclusion that the spin-orbit torques act mainly on the interfacial CrSBr layer and show that the linewidth modulation effects can be switched by changing the initial sublattice orientations.

\vspace{0.25cm}
\subsection{Damping-like spin-orbit torque efficiency calculation}
\noindent For device 1 analyzed in the main text, we estimate the lateral width of the PtTe$_2$ channel in the device region from the optical microscope image (main text Fig.\ 1A) as W = 3.07(9) $\rm{\mu}$m. The thicknesses of the PtTe$_2$ and CrSBr layers are measured using atomic-force microscopy, t$_{nm}$ = 93(2) nm, t$_{CrSBr}$ = 1.5(4) nm. The parameters for device 2 are W = 2.7(2) $\rm{\mu}$m, t$_{nm}$ = 61.1(6) nm and t$_{CrSBr}$ = 1.3(1) nm.

\vspace{0.5cm}
\noindent We estimate the saturation magnetization of our bilayer CrSBr using values reported in the literature. Firstly, we deduce a magnetic moment of ~1.95 $\rm{\mu_B}$ per Cr atom at 85 K for bulk CrSBr, by interpolating data from the literature (from Fig.\ 4f of ref.\ \cite{lee2021magnetic}). We calculate a unit cell volume of 132.11 Å$^3$ from the lattice parameters a = 3.5066(1) Å, b = 4.7485(1) Å, and c = 7.9341(2) Å\cite{lee2021magnetic}. Then, given 2 Cr atoms per unit cell, we get a saturation magnetization value of $\rm{\mu_0 M_{s, bulk}}$ = 0.35 T. Recent NV magnetometry measurements report a reduction in magnetization from bulk to bilayer CrSBr of around $\approx 20\%$ for encapsulated flakes\cite{tschudin2023nanoscale}, from which we estimate for our bilayer at 85 K that $\rm{\mu_0 M_{s, 2L}}$ = 0.28 T (Fig.\ \ref{FigureS3}D). The uncertainty in this value is at least 10's of percent.

\vspace{0.5cm}
\noindent Using these parameters and cosine fits to the dc bias induced linewidth modulation, Eq.\ \ref{eqn:AFMRDcbiasXi} gives a damping-like efficiency of $ \rm{\xi^{device 1}_{DL}}$ = 0.29(2) and $\rm{\xi^{device 2}_{DL}} = 0.39(4)$ for devices 1 and 2 respectively.

\subsection{Effect on the fitted linewidth modulation of variations in the magnetization angles as a function of magnetic-field magnitude.}
\noindent As we vary the magnetic field at fixed frequency, the magnetization angle $\theta_1$ evolves as we sweep through a resonance.  We have analyzed to what extent this evolution may affect the fitting of our linewidths and our spin-orbit torque analysis.  

\vspace{0.5cm}
\noindent For the data we report, we calculate an expected $\theta_1$ variation of $\Delta\theta_1 \approx \pm 9\degree$ as the applied magnetic field is swept through our broadest resonances (the error bars for the values of $\theta_1$ in Fig.\ 4C of the main text reflect this variation). The consequence of this variation is small around $\theta_1 = 0\degree$ ($\Delta\cos{\theta_1} \approx -2\%$), and is largest around $\theta_1 = 90\degree$ ($\Delta\cos{\theta_1} \approx \pm15\%$). We have also performed LLGS simulations with different intrinsic linewidths ($\alpha = 0.033$ and $\alpha = 0.066$) and fitted the numerically-calculated V$\rm{_{mix}}$ resonance peaks over different field ranges ($\Delta\theta_1 \approx \pm4.5\degree$ and $\Delta\theta_1 \approx \pm9\degree$). As shown in Fig.\ \ref{FigureS11}, linewidth modulation values from the simulated fits show only slight deviations of less than 5$\%$ from the analytical trend of $\cos{\theta_1}$. 
These estimated errors are within the range of our experimental uncertainties and don't significantly change our spin-orbit torque efficiency calculations or our findings that the spin-orbit torque acts selectively on one spin sublattice.

\newpage
\null
\vfill
\vfill

\vfill
\vfill

\vfill
\vfill

\vfill
\vfill

\vfill
\vfill

\vfill
\vfill

\vfill
\vfill

\vfill
\vfill

\vfill
\vfill

\vfill
\begin{figure}[htpb]
    \centering
    \includegraphics[width=\textwidth]{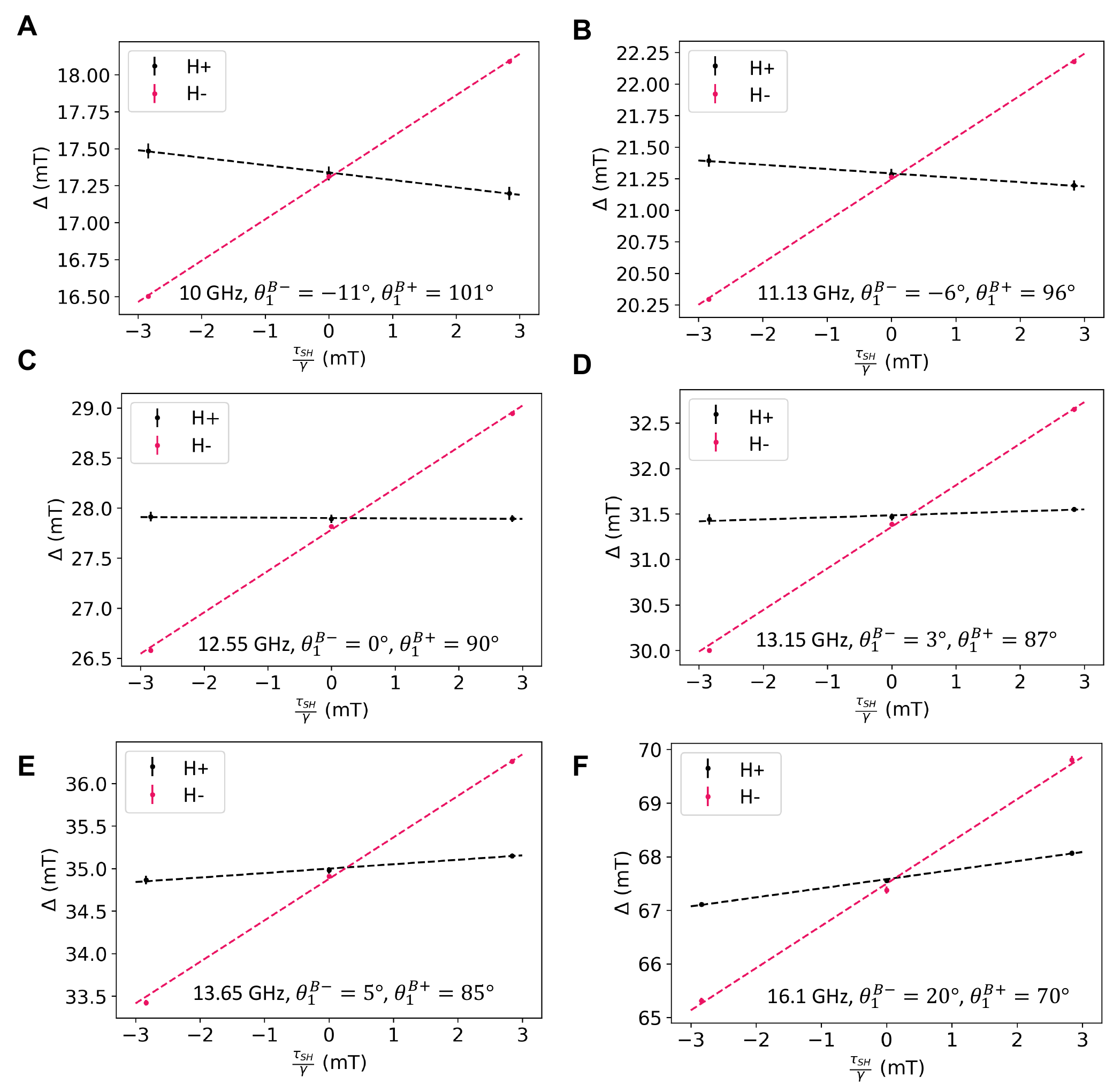}
    \caption{\textbf{Simulated linewidth vs spin-orbit torque from numerical integration of the coupled LLGS equations.} (\textbf{A}-\textbf{F}) Linear fits to the simulated linewidth modulation at various I$\rm{_{rf}}$ frequencies corresponding to different $\rm{\theta_1}$. The spin-orbit torque is proportional to the applied dc-bias current I$_{dc}$ and torque efficiency $\xi_{SH}$\cite{liu2011spin, karimeddiny2021resolving} (S.I.\ sections 2.3, 2.5). Slopes extracted from the linear fits show a 
    cos$\rm{\theta_1}$ dependence (Main text Fig.\ 4D).}
    \label{FigureS10} 
\end{figure}
\vfill

\vfill
\begin{figure}[htpb]
    \centering
    \includegraphics[width=\textwidth]{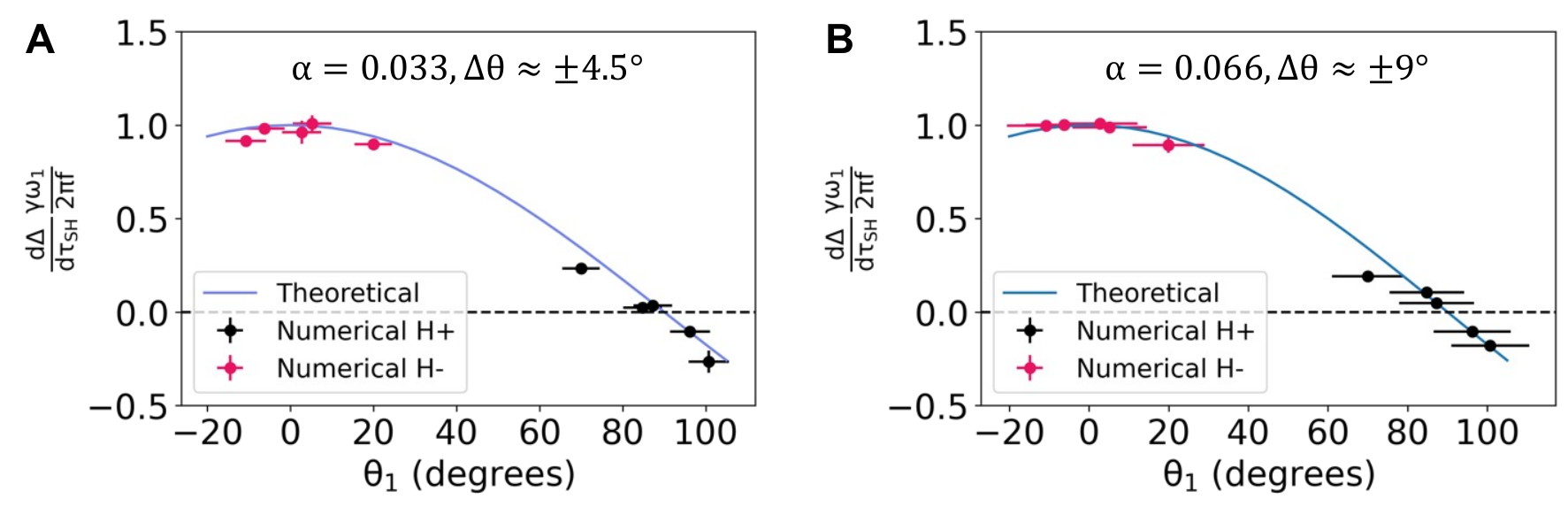}
    \caption{\textbf{Simulated linewidth modulation for two different intrinsic linewidths.} (\textbf{A}) Linewidth modulation for a system with $\alpha = 0.33$, from resonance fits over a range $\Delta\theta \approx \pm4.5\degree$. (\textbf{B}) Fitted modulation for $\alpha = 0.66$ over a range $\Delta\theta \approx \pm9\degree$. Blue curves indicate the predicted theoretical $\cos{\theta_1}$ angle dependence.
    }\label{FigureS11}
\end{figure}
\vfill

\vfill
\begin{figure}[htpb]
    \centering
    \includegraphics[width=\textwidth]{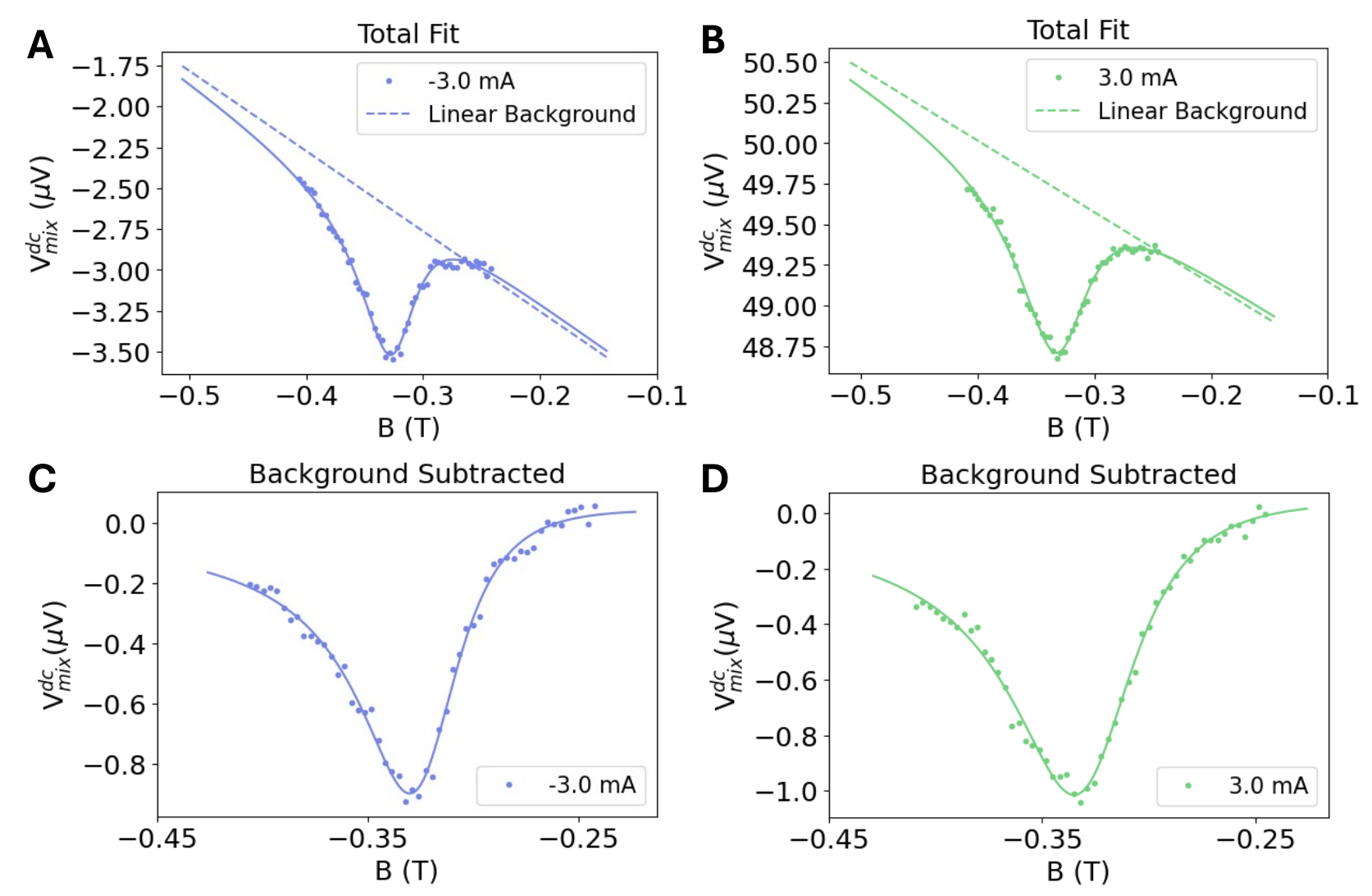}
    \caption{\textbf{Subtraction of non-resonant linear background.} (\textbf{A},\textbf{B}) ST-AFMR $V^{dc}_{mix}$ fitted with Lorentzian components and a non-resonant linear background at positive and negative dc current biases. (\textbf{C}, \textbf{D}) ST-AFMR $V^{dc}_{mix}$ with the non-resonant linear background subtracted.
    }\label{FigureS12}
\end{figure}
\vfill

\clearpage
% \bibliography{scibib}
% \bibliographystyle{science}